\documentstyle[12pt,epsf]{article}
\makeatletter

\@addtoreset{equation}{section}
\makeatother
\begin{document}
\baselineskip 0.8cm

\newcommand{\gsim}{ \mathop{}_{\textstyle \sim}^{\textstyle >} }
\newcommand{\lsim}{ \mathop{}_{\textstyle \sim}^{\textstyle <} }
\newcommand{\vev}[1]{ \left\langle {#1} \right\rangle }
\newcommand{\EV}{ {\rm eV} }
\newcommand{\KEV}{ {\rm keV} }
\newcommand{\MEV}{ {\rm MeV} }
\newcommand{\GEV}{ {\rm GeV} }
\newcommand{\TEV}{ {\rm TeV} }
\newcommand{\DS}{\displaystyle}
\def\tr{\mathop{\rm tr}\nolimits}
\def\Tr{\mathop{\rm Tr}\nolimits}
\def\Re{\mathop{\rm Re}\nolimits}
\def\Im{\mathop{\rm Im}\nolimits}
\def\simgt{\mathop{>}\limits_{\displaystyle{\sim}}}
\def\simlt{\mathop{<}\limits_{\displaystyle{\sim}}}
\setcounter{footnote}{0}

\begin{titlepage}

\begin{flushright}
UT-ICEPP 03-04\\
KEK-TH-888\\
\end{flushright}

\vskip 2cm
\begin{center}
{\large \bf  Probing the CP nature of the Higgs bosons \\
by $t\overline{t}$ production at photon linear colliders}
\vskip 1.2cm
Eri Asakawa$^{a, b}$, and Kaoru Hagiwara$^{b}$

\vskip 0.4cm

$^{a}$ {\it International Center for Elementary Particle Physics, \\
        University of Tokyo, Tokyo 113-0033, Japan}\\
$^{b}$ {\it Theory Group, KEK, Tsukuba, Ibaraki 305-0801, Japan}
\vskip 1.5cm

\abstract{
We study effects of heavy Higgs bosons on the $t \overline{t}$
production process at photon linear colliders.
The interference patterns between the resonant Higgs-production
amplitudes and the continuum QED amplitudes are examined.
The patterns tell us not only the
CP nature of the Higgs bosons but also the phase of the 
$\gamma\gamma$--Higgs vertex 
which gives new information about the Higgs couplings 
to new charged particles.
We point out that it is necessary to use circularly polarized
photon beams to produce efficiently heavy Higgs bosons whose
masses exceed the electron beam energy, and show that the above
interference patterns of the production amplitudes can be studied
by observing
$t$ and $\overline{t}$ decay angular distributions.
Analytic expressions for the helicity amplitudes
for the sequential process $\gamma\gamma \rightarrow t
\overline{t} \rightarrow (bW^+) (\overline{b}W^-) \rightarrow
(b f_1 \overline{f}_2)  (\overline{b} f_3 \overline{f}_4)$ are
presented in terms of the generic $\gamma\gamma \rightarrow
t \overline{t}$ production amplitudes.
}

\end{center}
\end{titlepage}

\renewcommand{\thefootnote}{\arabic{footnote}}
\setcounter{footnote}{0}

%
%
%
%
\section{Introduction}

The scalar sector of
the Standard Model (SM) consists of one $SU(2)_{\rm w}^{}$ doublet.
After the electroweak symmetry breaking (EWSB),
a neutral CP-even Higgs boson remains as a physical state.
Although the SM is consistent with the current experimental
data, 
new physics will be indispensable
if we consider the hierarchy
between the electroweak scale and the Planck scale,
a failure of the gauge coupling unification etc.
as serious problems.
It is natural that new physics modify the mechanism
of the EWSB. Such modification may lead to appearance of
Higgs bosons with various CP properties.
In the case that an extra doublet extends
the scalar sector of the SM, extra two neutral and
two charged Higgs bosons should be observed.
If CP is a good symmetry of the scalar sector, 
one additional neutral boson is CP-even and the other is CP-odd.
Therefore, probing the CP property as well as the
masses, the decay widths and the couplings of all the Higgs bosons 
is  necessary for exploring the Higgs sector.

One of colliders which can play an important role in
studying the Higgs sector is a photon linear collider (PLC),
an option of $e^+ e^-$ linear colliders~\cite{JLC,TESLA,NLC}.
The energy of the colliding photons, which are
obtained by the backward Compton scattering of laser light
on high-energy electrons, reaches about $80$\% of 
the energy of the original electron beam~\cite{polarization}. 
Since neutral Higgs bosons are produced 
as $s$-channel resonances via loops of charged massive particles,
we can detect the Higgs bosons whose masses are less than
about 80\% of the collision energy of a parent $e^+ e^-$
collider. Thus, a PLC has a great advantage of 
detecting heavy neutral Higgs bosons 
whose masses exceed the reach of the LHC and an $e^+ e^-$ LC
especially for those of the minimal supersymmetric
SM (MSSM)~\cite{MM}.
For light Higgs bosons, it has been well known that
the $\gamma\gamma$ decay widths of the Higgs
bosons can be accurately measured~\cite{ggwidth}. 
The measurement is important because the contribution from
heavy charged particles which couple to the Higgs bosons
does not decouple from the vertex if their masses originate
from the EWSB.
As for CP nature of Higgs bosons,
CP-even and CP-odd Higgs bosons can be 
clearly distinguished by utilizing the linear polarization 
of colliding photons~\cite{Gunion}. This powerful
technique, however, is effective to
probe the CP nature of relatively light
Higgs bosons only, because the linear polarization transfer
of the Compton back-scattered laser light decreases 
significantly when the photon energy is more than
half the electron beam energy~\cite{polarization, DESYtalk}.
For the heavier Higgs bosons whose masses exceed the electron
beam energy, $t\overline{t}$ production
process with circularly polarized photons is useful
to study their CP properties~\cite{AKSW, ACHL, GRS}. 

In this paper, we revisit the study of the CP nature of neutral Higgs
bosons through the $t \overline{t}$ production process at a PLC.
Such study has been performed in \cite{AKSW}, \cite{ACHL} 
and \cite{GRS}.
It has been shown in Sec.~4.4 of \cite{AKSW} that;
if we observe
sizable interference between the Higgs-resonant and QED-continuum
amplitudes for the two helicity combinations of the
top pairs produced
by circularly polarized colliding photons,
we can determine the CP parity of the Higgs bosons.
In \cite{ACHL}, the observables which are useful for
complete determination of
the $\gamma\gamma$-Higgs and $t\overline{t}$-Higgs
couplings have been presented,
in the presence of CP non-conserving interactions.
The accuracy of the determination of those couplings
has been studied in \cite{GRS}, by using the combined asymmetries
involving the circular polarization of colliding photons
and the charge of charged leptons in top decays with
a cut off on the lepton angle.

In this paper, we extend the study of \cite{AKSW},
and study the interference patterns of the resonant and the
continuum amplitudes in more detail for the $\gamma\gamma \rightarrow
t \overline{t}$ process by using the circularly polarized colliding
photons. We find that not only the squares of the helicity amplitudes
but also the real and imaginary parts of the interference between
the two helicity amplitudes can be measured by studying the angular
correlations of $t$ and $\overline{t}$ decay products
They are useful for deriving the information
on the CP nature of Higgs bosons.
It will also be shown that
these interference effects allow us to observe the complex phase of 
the $\gamma\gamma$-Higgs vertices. 

This paper is organized as follows. In Sect.~2 helicity amplitudes 
for the process $\gamma\gamma \rightarrow t \overline{t}$ are
given. In Sect.~3 observables which are sensitive to the CP-parity
of the Higgs bosons as well as the complex phase of the
$\gamma\gamma$-Higgs vertex
are discussed. Numerical estimates of
the observables which are introduced in Sect.~3 are 
performed in Sect.~4.
We give conclusions in the last section.
Analytic expressions for the helicity amplitudes
for the sequential process $\gamma\gamma \rightarrow t
\overline{t} \rightarrow (bW^+)(\overline{b}W^-) \rightarrow
(b f_1 \overline{f}_2)(\overline{b} f_3 \overline{f}_4$) are
presented in appendix A.

\section{Helicity amplitudes for the process $\gamma \gamma 
\rightarrow t \overline{t}$}

When the $\gamma\gamma$ collision energy reaches around 
mass of a spinless boson $\phi$ ($\phi=H$ or $A$ where $H$ and 
$A$ are the CP-even and CP-odd Higgs bosons respectively.),
the process 
\begin{eqnarray}
\gamma (k_1, \lambda_1) + \gamma (k_2, \lambda_2) \rightarrow
t (p, \sigma) + \overline{t} (\overline{p}, \overline{\sigma})
\end{eqnarray}
receives leading contributions from the diagrams 
in which the spinless 
boson is exchanged in the $s$-channel 
and the top quark is exchanged in the $t$- and $u$-channels.
The four-momenta and the helicities
of the participating particles in the colliding $\gamma \gamma$
center-of-mass frame are given in parentheses.
We adopt the notation~\cite{HZ1986} where the photon (fermion)
helicities are denoted by the signs in units of $\hbar$ ($\hbar$/2)
\footnote{For fermion helicities we often use the notation
$L$ and $R$ instead of $-$ and $+$}.
The helicity amplitudes of the process can be expressed as
\begin{eqnarray}
\label{ampeq}
{\cal M}_{\lambda_1 \lambda_2}^{\sigma \overline{\sigma}} =
\left[ {\cal M}_\phi \right]
_{\lambda_1 \lambda_2}^{\sigma \overline{\sigma}} +
\left[ {\cal M}_t \right]
_{\lambda_1 \lambda_2}^{\sigma \overline{\sigma}} ,
\end{eqnarray}
where the first term ${\cal M}_\phi$ stands for the $s$-channel
$\phi$-exchange amplitudes and the latter term ${\cal M}_t$
stands for the $t$- and $u$-channel top-quark-exchange amplitudes.
The resonant helicity amplitudes are calculated by using
the lowest-dimensional effective
Lagrangian of the form
\begin{eqnarray}
{\cal L}_{\phi \gamma \gamma} &=& \frac{1}{m_\phi}
\left( b_\gamma^H A_{\mu\nu} A^{\mu\nu} +
b_{\gamma}^A \widetilde{A}_{\mu\nu} A^{\mu\nu} \right) \phi ,
\\
{\cal L}_{\phi t \overline{t}} &=&
\overline{t} \left( d_t^H + i d_t^A \gamma_5 \right) t \phi,
\end{eqnarray}
where $A_{\mu \nu}= \partial_\mu A_\nu - \partial_\nu A_\mu$
and $\widetilde{A}_{\mu \nu} = \frac{1}{2}
\epsilon_{\mu \nu \rho \sigma}
A^{\rho \sigma}$ (where $\epsilon_{0123}=1$) are the photon field
strength tensor and its dual tensor, respectively.
The resonant amplitudes are then expressed as products
of the $\gamma \gamma \phi$
vertex function $A_{\phi}^{\lambda_1 \lambda_2}$,
the Higgs propagator factor $B_{\phi}$ and the decay vertex
$C_{\phi}^{\sigma \overline{\sigma}}$,
\begin{eqnarray}
\label{phieq}
\left[ {\cal M}_\phi \right]
_{\lambda_1 \lambda_2}^{\sigma \overline{\sigma}}
= A_\phi^{\lambda_1 \lambda_2} B_\phi C_\phi^{\sigma \overline{\sigma}}
\end{eqnarray}
where
\begin{eqnarray}
A_\phi^{\lambda_1 \lambda_2}= 
\left( b_\gamma^H + i \lambda_1 b_\gamma^A \right)
\frac{\hat{s}}{m_\phi}~ \delta_{\lambda_1 \lambda_2},
\end{eqnarray}
\begin{eqnarray}
B_\phi = \frac{1}{m_\phi^2-\hat{s}-i m_\phi \Gamma_\phi},
\end{eqnarray}
\begin{eqnarray}
C_\phi^{\sigma \overline{\sigma}}=
\left( \beta ~\sigma d_t^H -i d_t^A \right) 
\sqrt{\hat{s}}~\delta_{\sigma \overline{\sigma}}.
\end{eqnarray}
In the CP-conserving limit, the $H$- and $A$-exchange 
amplitudes are \cite{AKSW}
\begin{eqnarray}
\left[ {\cal M}_H \right]
_{\lambda_1 \lambda_2}^{\sigma \overline{\sigma}}
&= & \sigma~\beta~b_\gamma^H d_t^H~
\frac{\hat{s}}{m_H^2-\hat{s}-im_H^{} \Gamma_H}~
\frac{\sqrt{\hat{s}}}{m_H^{ }}~
\delta_{\lambda_1, \lambda_2}
\delta_{\sigma, \overline{\sigma}} , \label{Heq}\\
\left[ {\cal M}_A \right]
_{\lambda_1 \lambda_2}^{\sigma \overline{\sigma}}
&= & \lambda_1~~b_\gamma^A ~d_t^A~
\frac{\hat{s}}{m_A^2-\hat{s}-im_A^{} \Gamma_A}~
\frac{\sqrt{\hat{s}}}{m_A^{ }}~
\delta_{\lambda_1, \lambda_2}
\delta_{\sigma, \overline{\sigma}} , \label{Aeq}
\end{eqnarray}
where 
$\beta$ is the velocity of the top quarks and
$\hat{s}$ is the total energy-squared in the rest frame of 
$\gamma \gamma$ collisions.
The masses and the total decay widths
of the Higgs bosons are denoted by $m_{\phi}$ and $\Gamma_{\phi}$.

In the following, we sometimes use the predictions of the MSSM
as examples. The effective couplings are expressed in the MSSM as
\begin{eqnarray}
d_t^H&=&-\frac{gm_t}{2m_W}\frac{\sin\alpha}{\sin\beta},
\\ \nonumber
d_t^A&=&\frac{gm_t}{2m_W}\cot\beta,
\end{eqnarray}
for the $ttH$ and $ttA$ couplings, where $g$ is the SU(2) gauge
coupling, $\tan\beta=\langle v_u \rangle / \langle v_d \rangle$ 
is the ratio of the two Higgs 
vacuum expectation values, and $\alpha$ is the mixing angle between
the neutral real components of the two Higgs doublets and 
the two CP-even Higgs bosons.
The $\gamma \gamma H$ and $\gamma \gamma A$ couplings are
induced in the one loop level:
\begin{eqnarray}
b_\gamma^H(\hat{s})&=&\frac{\alpha g}{8\pi} \frac{m_H^{ }}{m_W}
\sum_i I_H^i \left( \frac{\hat{s}}{m_i^2} \right),
\\ \nonumber
b_\gamma^A(\hat{s})&=&- \frac{\alpha g}{8\pi} \frac{m_A^{ }}{m_W}
\sum_i I_A^i \left( \frac{\hat{s}}{m_i^2} \right).
\end{eqnarray}
The dimensionless loop functions $I_H^i$ and $I_A^i$ for all
the MSSM diagrams (labeled by the index $i$, where the masses of 
particles in the loops are expressed by $m_i$)
are found e.g. in \cite{HHG}. As long as the SUSY particles are
heavier than the top quark, the top quark contribution dominates
over all the other contributions. The effective couplings 
$b_\gamma^H$ and $b_\gamma^A$ are real when all the particles
in the loops are heavy, and become complex above the thresholds. 

The irreducible background to the resonant $\phi$-production
process is the non-resonant top-quark-exchange processes, whose
amplitudes are expressed in the tree level of QED as \cite{AKSW}
\begin{eqnarray}
\label{treeeq}
\left[ {\cal M}_t \right]
_{\lambda_1 \lambda_2}^{\sigma \overline{\sigma}}
&=& \frac{8 \pi \alpha Q_t^2}{1-\beta^2 \cos^2 \Theta} \times
\\ \nonumber
&~&
\huge\{ (\beta \sigma + \lambda_1) /\gamma
~\delta_{\lambda_1, \lambda_2} \delta_{\sigma, \overline{\sigma}}
- \beta /\gamma ~\sigma \sin^2 \Theta~
\delta_{\lambda_1, -\lambda_2} \delta_{\sigma, \overline{\sigma}}
\\ \nonumber
&~& - \beta (\sigma \lambda_1 + \cos\Theta) \sin\Theta~
\delta_{\lambda_1, -\lambda_2} \delta_{\sigma, -\overline{\sigma}}
\huge\}.
\end{eqnarray}
Here $1/\gamma = \sqrt{1-\beta^2} = 2m_t/\sqrt{\hat{s}}$
and $\Theta$ is the polar angle of the top-quark momentum
in the colliding $\gamma \gamma$ c.o.m.~frame.
In Table 1, the amplitudes in units of the common factor
$8\pi\alpha Q_t^2/(1-\beta^2 \cos^2 \Theta)$
are summarized. In the table, the photon helicities
$\lambda_1\lambda_2$ are given in the first column, and
the $t \overline{t}$ helicities $\sigma \overline{\sigma}$ 
are denoted as $RR$, $LL$, $RL$, $LR$ for
$(\sigma\overline{\sigma})=(++),~(--),~(+-),~(-+)$, respectively,
in the first row.
It should be noted that the four amplitudes in the
left top column of Table~\ref{treehelamp}, those for 
$\lambda_1=\lambda_2$ and $\sigma=\overline{\sigma}$,
interfere with the resonant amplitudes of 
eq.~(\ref{phieq}).
Furthermore, at high energies ($\beta\rightarrow 1$, $\gamma\gg 1$)
all the $\sigma=\overline{\sigma}$ amplitudes are suppressed by 
$1/\gamma$, among which the amplitudes for $\sigma=\overline{\sigma}=
-\lambda=-\overline{\lambda}$ are suppressed by $1/\gamma^3$.
These properties as well as the relative signs of 
the top-quark-exchange
amplitudes will be found useful in probing the CP nature of the
Higgs bosons in the following sections.  
\begin{table}[h]
\begin{center}
\caption[Tree helicity amplitudes]{\small
 The tree-level helicity amplitudes of $\gamma\gamma$ $\rightarrow
t\overline{t}$,
$\left[{\cal M}_t \right]_{\lambda_1 \lambda_2}
^{\sigma \overline{\sigma}}$, in eq.~(\ref{treeeq}).
 The common factor $8 \pi \alpha Q_t^2/(1-\beta^2 \cos^2\Theta)$ is
omitted in the table. The two photon helicities $\lambda_1 \lambda_2$
are given in the first column, and
the $t \overline{t}$ helicities $\sigma \overline{\sigma}$ 
are denoted as $RR$, $LL$, $RL$, $LR$ for
$(\sigma\overline{\sigma})=(++),~(--),~(+-),~(-+)$, respectively,
in the first row.
\label{treehelamp} }
\vspace{7mm}
\begin{tabular}{|c||c|c|c|c|}
\hline
 & $RR$ & $LL$ & $RL$ & $LR$ \\
\hline \hline
$++$ & $(1+\beta)/\gamma$ & $(1-\beta)/\gamma$ & $0$ & $0$ \\
\hline
$--$ & $-(1-\beta)/\gamma$ & $-(1+\beta)/\gamma$ & $0$ & $0$ \\
\hline
$+-$ & $-\beta/\gamma~\sin^2\Theta$ 
& $\beta/\gamma~\sin^2\Theta $ &
$ -\beta\sin\Theta(1+\cos\Theta) $ &
$ \beta\sin\Theta(1-\cos\Theta) $ \\
\hline
$-+$ & $-\beta/\gamma~\sin^2\Theta$ 
& $\beta/\gamma~\sin^2\Theta $ &
$ \beta\sin\Theta(1-\cos\Theta) $ &
$ -\beta\sin\Theta(1+\cos\Theta) $ \\
\hline
\end{tabular}
\label{tree}
\end{center}
\end{table}

\section{Determining the CP parity of the Higgs bosons}

\subsection{Overview}

The helicity dependence of the amplitudes discussed in the 
previous section is summarized in Table~\ref{helamp}.
We note here that the individual ($H$-exchange, $A$-exchange, and
$t$-exchange) amplitudes for the helicities 
$\lambda=\overline{\lambda}=-$ and $\lambda=\overline{\lambda}=
-\sigma=-\overline{\sigma}$ are obtained from 
the $\lambda=\overline{\lambda}=\sigma=\overline{\sigma}=+$
amplitudes $\left[ {\cal M}_{H,A,t} \right]^{++}_{++}$
by multipling the appropriate sign-factor representing the
CP transformation property and the kinematical factor
for the top-quark-exchange amplitudes. Here
$\left[ {\cal M}_{H,A,t}\right]^{++}_{++}$ are denoted by
${\cal M}_{H,A,t}$ for simplicity.
\begin{table}[h]
\caption{The helicity dependence of the amplitudes of
$\gamma(\lambda) \gamma(\lambda) \rightarrow 
t(\sigma) \bar{t}(\sigma)$, 
$[{\cal M}]_{\lambda \lambda}^{\sigma \sigma}$.
We denote $[{\cal M}]_{++}^{RR}$ as ${\cal M}$ for
${\cal M}_t$, ${\cal M}_H$ and ${\cal M}_A$, which denote the
top-, $H$- and $A$-exchange amplitudes, respectively.
The two photon helicities $\lambda \lambda$
are given in the first column, and
the $t \overline{t}$ helicities $\sigma \sigma$ 
are denoted as $RR$, $LL$ for
$(\sigma\overline{\sigma})=(++),~(--)$, respectively,
in the first row.
\label{helamp}}
\vspace{0.4cm}
\begin{center}
\begin{tabular}{|c||c|c|}
\hline
 & RR & LL \\
\hline \hline
 $++$ &
\begin{minipage}{1.0in}
\begin{center}
$~~~~~~{\cal M}_t $\\
$~~~~~~{\cal M}_H $\\
$~~~~~~{\cal M}_A $
\end{center}
\end{minipage}
&
\begin{minipage}{1.0in}
\begin{center}
$\frac{1-\beta}{1+\beta} {\cal M}_t $\\
$~-{\cal M}_H $\\
$~~~~{\cal M}_A$
\end{center}
\end{minipage}
\\
\hline
$--$&
\begin{minipage}{1.0in}
\begin{center}
$-\frac{1-\beta}{1+\beta}{\cal M}_t$
$~~~~~~~{\cal M}_H $\\
$~~~-{\cal M}_A$
\end{center}
\end{minipage}
&
\begin{minipage}{1.0in}
\begin{center}
$~-{\cal M}_t $\\
$~-{\cal M}_H $\\
$~-{\cal M}_A$
\end{center}
\end{minipage}
\\
\hline
\end{tabular}
\end{center}
\end{table}
When the polarization of the colliding beams is fixed,
e.g. as $\lambda=\overline{\lambda}=+$,
the sign of the $H$-production amplitude
changes when the helicities of final top pairs are flipped.
On the other hand,
the sign of the $A$-production amplitude
does not depend on the helicities of final top pairs.
The sign of the top-quark-exchange amplitudes does not depend on
the $t\overline{t}$ helicities, just like the $A$-exchange
amplitudes, but the amplitude is reduced  by a factor of
$(1-\beta)/(1+\beta)=1/[\gamma^2(1+\beta)^2]$ when the 
top-quark-helicity is opposite to the photon helicity,
$\lambda_1=\lambda_2=-\sigma=-\overline{\sigma}$.
Therefore, the top-quark-helicity dependence of the interference pattern
between the resonant amplitudes and the top-quark-exchange amplitudes
can be used to determine the CP parity of the Higgs resonance
\cite{AKSW}. 
It should further be noted that within the given helicity amplitude
the interference pattern below and above the resonance is also
a good probe the CP parity. In our phase connection,
${\cal M}_t$ is positive
at all $\hat{s}$, whereas the $\phi$-exchange amplitude
${\cal M}_\phi$ is
positive at low $\hat{s}$ where the absorptive part of the
$\phi\gamma\gamma$ vertex can be neglected for the dominant
top-quark loop contribution.
We should hence expect constructive interference
below the resonance when $\lambda_1=\lambda_2
=\sigma=\overline{\sigma}$. The above statements are valid for
both $H$ and $A$, or their arbitrary mixture when CP is violated.
The interference pattern for the $\lambda_1=\lambda_2
=-\sigma=-\overline{\sigma}$ amplitude
is expected to reverse for $H$, whereas it remains the same for
$A$. Both signs are possible when the resonance $\phi$ does not 
have a definite CP parity. 

Based on the above observation, we study carefully the 
interference patterns between the helicity amplitudes, 
that receive contribution from
the $s$-channel spin-0 resonance production.
In general, four types of observables can be studied in the
process $\gamma\gamma \rightarrow t\overline{t}$ where the
initial photon polarization can be controlled by 
the backward Compton scattering 
of the laser light and the $t\overline{t}$
polarization are measured through the angular distributions of
the correlated cascade decays, $t\rightarrow bW^+ \rightarrow 
b f_1 \overline{f}_2$ and $\overline{t} \rightarrow \overline{b}W^-
\rightarrow \overline{b}f_3 \overline{f}_4$.
All the observables which are sensitive to the spin-0
resonance contributions are listed below;
\begin{eqnarray}
&\bullet&~~|{\cal M}_{\lambda \lambda}^{RR}|^2,~~
|{\cal M}_{\lambda \lambda}^{LL}|^2~~~~~~~~~~~~~~~~~~
~~~~~~~~~~~~~~~~~~ {\rm for}~~\lambda=+,-; \label{A}
\\ 
&\bullet&~~\Re,\Im[{\cal M}_{\lambda \lambda}^{RR}
\left({\cal M}_{\lambda \lambda}^{LL}\right)^*]~~~~~~~~~~~
~~~~~~~~~~~~~~~~~~~ {\rm for}~~\lambda=+,-; \label{B}~~
\\ 
&\bullet&~~\Re,\Im[{\cal M}_{++}^{\sigma \sigma}
\left({\cal M}_{--}^{\sigma \sigma}\right)^*]~~~~~~~~~~~
~~~~~~~~~~~~~~~~~~ {\rm for}~~\sigma=R,L; \label{C}
\\
&\bullet&~~\Re,\Im[{\cal M}_{++}^{\sigma \sigma}
\left({\cal M}_{--}^{-\sigma, -\sigma}\right)^*]~~~~~~~~~~~
~~~~~~~~~~~~~~~~ {\rm for}~~\sigma=R,L. \label{D}
\end{eqnarray}
The observables (\ref{A}) have been studied in \cite{AKSW} and
they are found to be useful in distinguishing $A$ from $H$.
The observables (\ref{C}) have been studied in \cite{ACHL} and
are found to be effective in probing the CP nature of the neutral
Higgs sector, including the case of CP-violation. Unfortunately,
the observables (\ref{C}) require
linear polarization of the colliding
photon beams, whose magnitude is small for $z \equiv 
\sqrt{\hat{s}}/\sqrt{s} \gsim 0.5$ where $\sqrt{s}$ is the c.o.m.~energy
of a parent $e^- e^-$ collider~\cite{polarization, DESYtalk}.
In this article,
we concentrate on the observables (\ref{A}) and (\ref{B}), 
which can take
advantage of the high $\gamma\gamma$ luminosity at large
$z$ with high level of 
monochromaticity, that are obtained from the backward
Compton scattering
of circularly polarized laser lights on longitudinally polarized
electron beams. The CP-violating cases will be studied elsewhere
\cite{AHfuture}.
To our knowledge, the observables of the type (\ref{D}),
whose observation requires both the linearly polarized photons
and the angular correlations of $t$ and $\overline{t}$ decays,
have not been studied. 

\subsection{Observables}

Because the top quark polarizations are 
measured through its decay angular distribution~\cite{HMW, lepton}, 
we study the cascade process 
\begin{eqnarray}
\gamma(k_1, \lambda_1) + \gamma(k_2, \lambda_2) &\rightarrow&
t(p,\sigma) + \overline{t}(\overline{p},\overline{\sigma})
\label{a} \\
&\rightarrow& b(p_b,L)~W^+(p_W,\Lambda)+\overline{b}(\overline{p}_b,R)
~W^-(\overline{p}_W,\overline{\Lambda})
\label{b} \\
&\rightarrow& b(p_b,L)~f_1(p_1,L)~\overline{f}_2(p_2,R)
+\overline{b}(\overline{p}_b,R)~f_3(p_3,L)~\overline{f}_4(p_4,R)
\nonumber\\
\label{c}
\end{eqnarray}
where we assume the SM amplitudes for the decays, and neglect
masses of all final fermions including $b$ and $\overline{b}$.
The helicity amplitudes for the full process (\ref{c}),
${\bf M}_{\lambda_1 \lambda_2}$, are
given in appendix A. The differential cross section for 
arbitrary initial photon helicities
\begin{eqnarray}
\label{comdiffcross}
\frac{d\hat{\sigma}_{\lambda_1 \lambda_2}}{d\cos\Theta~d\cos\theta
~d\phi~d\cos\overline{\theta}~d\overline{\phi}~d\cos\theta^*
~d\phi^*~d\cos\overline{\theta}^*~d\overline{\phi}^*}
\\ \nonumber
= \frac{3 \beta}{32\pi \hat{s}}
\left| {\bf M}_{\lambda_1 \lambda_2}
(\Theta; \theta,\phi; \overline{\theta},\overline{\phi} ;
\theta^*,\phi^*; \overline{\theta}^*,\overline{\phi}^*) \right|^2
\times B_{12} B_{34}
\end{eqnarray}
is readily obtained in the zero-width limit of the top quarks and
the $W$ bosons. Here $B_{12}$ is the branching fraction 
of $W^+ \rightarrow f_1 \overline{f}_2$ decays, 
and $B_{34}$ is that of $W^- \rightarrow f_3 \overline{f}_4$,
$\hat{s}=(k_1+k_2)^2$ is the total-energy
squared in the colliding $\gamma\gamma$ c.o.m.~system,
$\Theta$ is the polar angle of the top-quark momentum
in this frame measured from the direction of the photon beam with
the momentum $k_1$, $\theta$ and $\phi$ ($\overline{\theta}$ and 
$\overline{\phi}$) are the polar and azimuthal angles, respectively,
of the $W^+$ ($W^-$) momentum in the $t$ ($\overline{t}$) rest-frame.
The polar angles ($\theta$ and $\overline{\theta}$) are
measured from the top-quark momentum direction 
in the $\gamma\gamma$ c.o.m.~frame and the azimuthal angles 
($\phi$ and $\overline{\phi}$) are measured from the $\gamma\gamma
\rightarrow t\overline{t}$ scattering plane.
Here we choose the common polar axis and the $\phi=\overline{\phi}=0$
plane to describe the $t \rightarrow bW^+$ and $\overline{t} \rightarrow
\overline{b}W^-$ decays, so that our coordinate frame for
$\overline{t} \rightarrow \overline{b}W^-$ decays is obtained from the
frame used for $t \rightarrow bW^+$ decays by a single boost along
the top-quark momentum direction.
Finally, $\theta^*$ and $\phi^*$ ($\overline{\theta}^*$ and 
$\overline{\phi}^*$) are the polar and azimuthal angles, respectively,
of the $\overline{f_2}$ ($f_3$) momentum in the $W^+$ ($W^-$) rest-frame.
The polar angle $\theta^*$ ($\overline{\theta}^*$) is
measured from the $W^+$ ($W^-$) momentum direction in the
$t$ ($\overline{t}$) restframe, and the azimuthal angle 
$\phi^*$ ($\overline{\phi}^*$) is measured from the
$W^+b$ ($W^- \overline{b}$)
decay plane in the $\gamma\gamma$ collision c.o.m.~frame.
The origins of the azimuthal angles are chosen such that the y-axis
for $\phi=\overline{\phi}=\frac{\pi}{2}$ is along 
the $\vec{k_1} \times \vec{p}$ direction in the $\gamma\gamma$
c.o.m.~frame, that for $\phi^*=\frac{\pi}{2}$
($\overline{\phi}^*=\frac{\pi}{2}$) is along the
$\vec{p} \times \vec{p}_{W^+}$ 
($\vec{\overline{p}} \times \vec{p}_{W^-}$)
direction also in the $\gamma\gamma$ c.o.m.~frame. 

If we assume that the top-quark decays are essentially 
described by the SM amplitudes as above, it is straightforward
to extract all the four observables (\ref{A}) and (\ref{B}),
for a given initial photon polarization ($\lambda_1=\lambda_2=+$ or $-$),
by studying the $t$ and $\overline{t}$ decay angular distributions.
Optimal accuracy of such measurements can readily be estimated
by using the exclusive distributions \cite{NGHIKK} for a given
range of the scattering angle $\Theta$. Such measurements should be 
especially effective near the $t\overline{t}$ threshold where
the $\Theta$-dependence of the background amplitude is moderate.
In this article, we present a primitive version of such analysis
where we assume that the exclusive distributions of the 
$\gamma\gamma \rightarrow t\overline{t} \rightarrow
bW^+ \overline{b}W^-$
process (\ref{b}) are measured for transverse and longitudinally
polarized $W$'s ($W_T$ and $W_L$, respectively) separately.
We assume that the $W^+$ ($W^-$) helicity is measured in the
$t$ ($\overline{t}$) rest frame. Such distributions are in
principle  measurable when the $W$-pair decays hadronically or
semi-leptonically. When both $W$'s decay leptonically,
presence of two energetic neutrinos in the final state makes it
impossible to reconstruct the $W$ momenta uniquely.
It should further be noted that $W_T$ and $W_L$ can be distinguished
experimentally even when the $W$ decays hadronically,
though less efficiently than the leptonic-decay case.

The differential cross sections for polarized $W$'s are now 
expressed compactly as follows:
\begin{eqnarray}
\frac{d \hat{\sigma}_{\lambda_1 \lambda_2}
^{\Lambda \overline{\Lambda}}}{d\cos\Theta~d\cos\theta~d\phi~
d\cos\overline{\theta}~d\overline{\phi}} =
\frac{3 \beta}{32 \pi \hat{s}} \left|
{\bf M}_{\lambda_1 \lambda_2}^{\Lambda \overline{\Lambda}}
(\Theta; \theta,\phi,\overline{\theta},\overline{\phi})
\right|^2 .
\end{eqnarray}
Explicit forms of the helicity amplitude
${\bf M}_{\lambda_1 \lambda_2}^{\Lambda
\overline{\Lambda}}$ appear in appendix~B.
Here, we consider the case of $\lambda_1=\lambda_2=\lambda$,
because high luminosity and high degree of polarization for
energetic two photon pairs can be achieved at a PLC.
The four relevant squared matrix elements for 
$\lambda_1=\lambda_2=\lambda$ are
\begin{eqnarray}
\label{ggbwbwLL}
|{\bf M}_{\lambda \lambda}^{LL}|^2 = \frac{B_L^2}{16\pi^2}
&\{&
|{\cal M}_{\lambda \lambda}^{RR}|^2 (1+\cos\theta)
(1+\cos\overline{\theta})
\\ \nonumber
&+&
|{\cal M}_{\lambda \lambda}^{LL}|^2 (1-\cos\theta)
(1-\cos\overline{\theta})
\\ \nonumber
&+&2\Re \left[ {\cal M}_{\lambda \lambda}^{RR} \cdot 
\left( {\cal M}_{\lambda \lambda}^{LL} \right)^* \right]
\sin\theta \sin\overline{\theta} \cos(\phi-\overline{\phi})
\\ \nonumber
&-&2\Im \left[ {\cal M}_{\lambda \lambda}^{RR} \cdot 
\left( {\cal M}_{\lambda \lambda}^{LL} \right)^* \right]
\sin\theta \sin\overline{\theta} \sin(\phi-\overline{\phi})\},
\\
\label{ggbwbwLT}
|{\bf M}_{\lambda \lambda}^{LT}|^2 = \frac{B_L B_T}{16\pi^2} 
&\{&
|{\cal M}_{\lambda \lambda}^{RR}|^2 (1+\cos\theta)
(1-\cos\overline{\theta})
\\ \nonumber
&+&
|{\cal M}_{\lambda \lambda}^{LL}|^2 (1-\cos\theta)
(1+\cos\overline{\theta})
\\ \nonumber
&-&2\Re \left[ {\cal M}_{\lambda \lambda}^{RR} \cdot 
\left( {\cal M}_{\lambda \lambda}^{LL} \right)^* \right]
\sin\theta \sin\overline{\theta} \cos(\phi-\overline{\phi})
\\ \nonumber
&+&2\Im \left[ {\cal M}_{\lambda \lambda}^{RR} \cdot 
\left( {\cal M}_{\lambda \lambda}^{LL} \right)^* \right]
\sin\theta \sin\overline{\theta} \sin(\phi-\overline{\phi})\},
\\
\label{ggbwbwTL}
|{\bf M}_{\lambda \lambda}^{TL}|^2 = \frac{B_L B_T}{16\pi^2} 
&\{&
|{\cal M}_{\lambda \lambda}^{RR}|^2 (1-\cos\theta)
(1+\cos\overline{\theta})
\\ \nonumber
&+&
|{\cal M}_{\lambda \lambda}^{LL}|^2 (1+\cos\theta)
(1-\cos\overline{\theta})
\\ \nonumber
&-&2\Re \left[ {\cal M}_{\lambda \lambda}^{RR} \cdot 
\left( {\cal M}_{\lambda \lambda}^{LL} \right)^* \right]
\sin\theta \sin\overline{\theta} \cos(\phi-\overline{\phi})
\\ \nonumber
&+&2\Im \left[ {\cal M}_{\lambda \lambda}^{RR} \cdot 
\left( {\cal M}_{\lambda \lambda}^{LL} \right)^* \right]
\sin\theta \sin\overline{\theta} \sin(\phi-\overline{\phi})\},
\\
\label{ggbwbwTT}
|{\bf M}_{\lambda \lambda}^{TT}|^2 = \frac{B_T^2}{16\pi^2}
&\{&
|{\cal M}_{\lambda \lambda}^{RR}|^2 (1-\cos\theta)
(1-\cos\overline{\theta})
\\ \nonumber
&+&
|{\cal M}_{\lambda \lambda}^{LL}|^2 (1+\cos\theta)
(1+\cos\overline{\theta})
\\ \nonumber
&+&2\Re \left[ {\cal M}_{\lambda \lambda}^{RR} \cdot 
\left( {\cal M}_{\lambda \lambda}^{LL} \right)^* \right]
\sin\theta \sin\overline{\theta} \cos(\phi-\overline{\phi})
\\ \nonumber
&-&2\Im \left[ {\cal M}_{\lambda \lambda}^{RR} \cdot 
\left( {\cal M}_{\lambda \lambda}^{LL} \right)^* \right]
\sin\theta \sin\overline{\theta} \sin(\phi-\overline{\phi})\},
\end{eqnarray}
where $B_L=m_t^2/(m_t^2+2m_W^2)$ and $B_T=2m_W^2/(m_t^2+2m_W^2)$
are the branching ratios of the decays $t \rightarrow b W_L^+$
($\overline{t} \rightarrow \overline{b} W_L^-$) and 
$t \rightarrow b W_T^+$
($\overline{t} \rightarrow \overline{b} W_T^-$), respectively.
It is clear that $|{\cal M}_{\lambda \lambda}^{RR}|^2$ and
$|{\cal M}_{\lambda \lambda}^{LL}|^2$ are obtained by
integrating out the $\phi-\overline{\phi}$ azimuthal angle
distributions, and they can be distinguished by using the
$W^+$ and $W^-$ polar angle ($\theta$ and $\overline{\theta}$)
distributions. Since it is necessary to distinguish $\theta$
from $\overline{\theta}$ ($W^+$ from $W^-$), semi-leptonic
decay modes should be used for the discrimination.
$\Re[{\cal M}_{\lambda_1 \lambda_2}^{RR} \cdot 
\left( {\cal M}_{\lambda_1 \lambda_2}^{LL} \right)^*]$ and $\Im
[{\cal M}_{\lambda_1 \lambda_2}^{RR} \cdot 
\left( {\cal M}_{\lambda_1 \lambda_2}^{LL} \right)^*]$ are obtained simply by
projecting out the $\cos(\phi-\overline{\phi})$ and
$\sin(\phi-\overline{\phi})$ distributions. Both $\phi$ and
$\overline{\phi}$ are observable when the $W^+ W^-$ pair decays
semi-leptonically. Because the above four distributions can be
measured independently, consistency among the four measurements 
can be checked.

We note here that the cross section for $\lambda_1=\lambda_2=\lambda$
without observing the $W$ polarization
can be written compactly as follows:
\begin{eqnarray}
\label{ggbwbw}
&&\frac{d \hat{\sigma}_{\lambda \lambda}}
{
d\cos\Theta~d\cos\theta~d\cos\overline{\theta}~d\phi~d\overline{\phi}}
= \frac{3 \beta}{32 \pi \hat{s}} \times \frac{1}{16\pi^2} \times
\\ \nonumber
&&~~~\{ |{\cal M}_{\lambda \lambda}^{RR}|^2 
\left[(B_L^2+B_T^2)(1+\cos\theta \cos\overline{\theta})
     +2 B_L B_T(1-\cos\theta \cos\overline{\theta})
     +(B_L^2-B_T^2)(\cos\theta + \cos\overline{\theta}) \right]
\\ \nonumber
&&~~~+ |{\cal M}_{\lambda \lambda}^{LL}|^2 
\left[(B_L^2+B_T^2)(1+\cos\theta \cos\overline{\theta})
     +2 B_L B_T(1-\cos\theta \cos\overline{\theta})
     -(B_L^2-B_T^2)(\cos\theta + \cos\overline{\theta}) \right]
\\ \nonumber
&&~~~+ 2 \Re \left[{\cal M}_{\lambda \lambda}^{RR} 
 ({\cal M}_{\lambda \lambda}^{LL})^*\right]
\left[(B_L-B_T)^2 
\sin\theta \sin\overline{\theta} \cos(\phi-\overline{\phi})
\right]
\\ \nonumber
&&~~~+ 2 \Im \left[{\cal M}_{\lambda \lambda}^{RR} 
({\cal M}_{\lambda \lambda}^{LL})^*\right]
\left[-(B_L-B_T)^2 
\sin\theta \sin\overline{\theta} \sin(\phi-\overline{\phi})
\right] \}.
\end{eqnarray}
Because $B_L$ and $B_T$ have different numerical values,
$B_L \simeq 0.7$ and $B_T \simeq 0.3$,
we can obtain the
four observables $|{\cal M}_{\lambda \lambda}^{RR}|^2$,
$|{\cal M}_{\lambda \lambda}^{LL}|^2$, 
$\Re[{\cal M}_{\lambda \lambda}^{RR}
({\cal M}_{\lambda \lambda}^{LL})^*]$ and
$\Im[{\cal M}_{\lambda \lambda}^{RR}
({\cal M}_{\lambda \lambda}^{LL})^*]$
even without observing the $W$ polarization.
In the following discussions, we adopt the simple expression
eq.~(\ref{ggbwbw}) in order to avoid repeating similar equations
four times. It should be understood that the measurements can
be improved significantly by using the $W$ polarization information,
as shown in eqs. (\ref{ggbwbwLL}) to (\ref{ggbwbwTT}).

\section{Numerical Estimates}
\subsection{Convoluted cross sections with energy distribution of 
photon beams}

The Compton back-scattered photons have broad energy distribution
with the maximal value $E_\gamma^{max}=\frac{x}{x+1}E_e$
with $x \equiv 4E_e \omega_L/m_e^2$, in the zero angle limit
of the Compton scattering. $E_e$ and $\omega_L$ are the electron
and laser photon energy. The circularly polarized laser photons
and longitudinally polarized electrons help the broad distribution
to peak near the high-energy end point
where the colliding photons are highly polarized.

\begin{figure}[h]
\begin{center}
\epsfxsize=15cm
\epsfysize=10cm
\epsffile{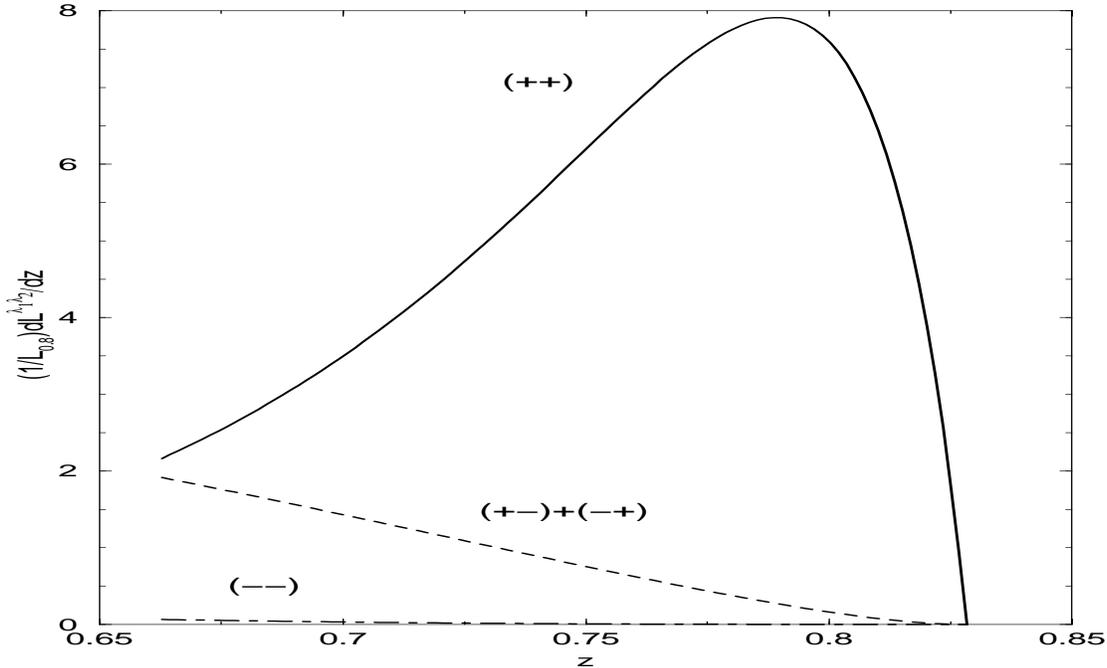}
\caption{The $\gamma \gamma$ liminosity functions 
normalized by ${\cal L}_{0.8}$,
the luminosity integrated over the region $z \geq 0.8z_m$
where $z=\sqrt{s}_{\gamma\gamma}/\sqrt{s}_{ee}$ and
$z_m=x/(x+1)$ is the maximum energy fraction.
The distributions of different $\gamma\gamma$ helicity
combinations, (++), ($--$), ($+-$) and ($-+$), are
shown separately for $P_l=-1.0$, $P_e=0.9$ and $x=4.8$.
}\label{luminosity}
\end{center}
\end{figure}

Fig.~\ref{luminosity} shows the $\gamma\gamma$ 
collision energy distribution 
which is calculated by the tree-level formula of the backward Compton 
scattering~\cite{polarization} for $x=4.8$
assuming complete polarization for laser photons 
($P_l=-1.0$) and 90\% polarization for electrons ($P_e=0.9$).
The distributions are shown for each combination
of $\gamma\gamma$ helicities.
The horizontal axis indicates the 
$\gamma\gamma$ collision energy ($\sqrt{\hat{s}}$) normalized by
the $ee$ c.o.m.~energy ($\sqrt{s}$), 
that is, $z=\sqrt{\hat{s}}/\sqrt{s}$.
The large $z$ region where the energy distribution is peaked
and dominated by the $++$ combination ($\lambda_1=\lambda_2=+$)
is most useful for the study
of $J_z=0$ mode in the $\gamma\gamma$ collision.
It is expected that the $\gamma\gamma$ luminosity in the region
$z \geq 0.8z_m=0.8\frac{x}{x+1}$ will account for about 10\% of
the geometric luminosity of electron-electron collisions,
 $L_{ee}^{geom}$ \cite{polarization}, 
\begin{eqnarray}
{\cal L}_{0.8} \equiv \sum_{\lambda_1,~\lambda_2}
\int_{0.8z_m}^{z_m} dz \frac{d{\cal L}^{\lambda_1 \lambda_2}}{dz}
\approx 0.1 {\cal L}_{ee}^{geom}.
\end{eqnarray}
In the lower energy region, $z \lsim 0.8 z_m$, both the spectrum
and the polarization receive significant non-linear corrections
so that the Compton scattering becomes a poor approximation.
We therefore normalized the $\gamma\gamma$ luminosity distributions
by ${\cal L}_{0.8}$ in Fig.~\ref{luminosity}.
All our convoluted cross sections are calculated for the 
$\gamma\gamma$ luminosity distributions normalized by 
${\cal L}_{0.8}$. The expected number of events is hence
obtained by multiplying the convoluted cross sections by
${\cal L}_{0.8} \approx 0.1{\cal L}_{ee}^{geom}$. Though
our luminosity functions based on Compton scattering are
not reliable at $z \lsim 0.8 z_m$ or $z \lsim 0.66$ for 
$x=4.8$, in this report we consider $t \overline{t}$ production
at a $\sqrt{s}_{ee}=500$ GeV collider, and hence our study is limited
to the region $z \geq 2m_t /\sqrt{s} \approx 0.7$.

Because of the above broad $\gamma\gamma$ energy distributions,
we cannot observe the $\gamma\gamma \rightarrow t\overline{t}$
production cross section at a given $\gamma\gamma$ energy,
$\sqrt{s}_{\gamma\gamma} \equiv \sqrt{\hat{s}}$. Instead we should
use the invariant mass of the final $t\overline{t}$ pair system,
$m_{t\overline{t}}$, as a measure of the colliding $\gamma\gamma$
energy. Although $m_{t\overline{t}}$ can in principle measured
event by event when a produced $t\overline{t}$ pair decays
hadronically or semi-leptonically, we should expect uncertainties
due to finite resolutions and non-Hermiticity of a detector.
We introduce a smearing function 
\begin{eqnarray}
G(\sqrt{\hat{s}}-m_{t\overline{t}},\Delta)
= \frac{1}{\sqrt{2\pi} \Delta} 
\exp\left[ -\frac{1}{2} \left( \frac{\sqrt{\hat{s}}-m_{t\overline{t}}}
{\Delta}\right)^2 \right],
\end{eqnarray}
between the true
$m_{t\overline{t}}=\sqrt{\hat{s}}$ and the $m_{t\overline{t}}$.
The observable cross sections can then be approximated as  
\begin{eqnarray}
\frac{d\sigma}{d m_{t\overline{t}}}
\equiv \int_{0}^{z_m \sqrt{s}} d \sqrt{\hat{s}}
\sum_{\lambda_1,~\lambda_2} \frac{1}{{\cal L}_{0.8}}
\frac{d {\cal L}^{\lambda_1 \lambda_2}}{d\sqrt{\hat{s}}}
\hat{\sigma}_{\lambda_1 \lambda_2}(\sqrt{\hat{s}})
~G(\sqrt{\hat{s}}-m_{t\overline{t}},\Delta).
\label{obscross}
\end{eqnarray}
When we set $\Delta=0$ GeV, the $m_{t\overline{t}}$ distributions
reproduce the $\sqrt{s}_{\gamma\gamma}$ distributions.

In eq.~(\ref{obscross}),
the $\gamma\gamma$ luminosity integrated over $z \geq 0.8z_m$ 
is denoted by
${\cal L}_{0.8}$ and the luminosity distribution for each
$\gamma\gamma$ helicity combination is expressed by 
$d{\cal L}^{\lambda_1 \lambda_2}/d \sqrt{\hat{s}}$.
Thus, the expected number of events with 
$m_{min}\leq m_{t\overline{t}}\leq m_{max}$
is estimated by the formula;
\begin{eqnarray}
N(m_{min}\leq m_{t\overline{t}} \leq m_{max}) 
= 0.1 L_{ee}^{geom} \times
\int_{m_{min}}^{m_{max}} dm_{t\overline{t}} 
\frac{d\sigma}{dm_{t\overline{t}}}.
\end{eqnarray}
It is notable that
the geometric $ee$ luminosity $L_{ee}^{geom}$
can be larger than the nominal $e^+ e^-$
luminosity $L_{ee}$. According to the TESLA design~\cite{TESLA},
\begin{eqnarray}
L_{ee}=3\times10^{34} {\rm cm}^{-2}{\rm s}^{-1},~~~
L_{ee}^{geom}=12\times10^{34} {\rm cm}^{-2}{\rm s}^{-1}
\end{eqnarray}
at $\sqrt{s}=500$ GeV have been reported.

\subsection{Results}

We consider the decay angular distribution of $t\overline{t}$
pairs produced via $\gamma\gamma$ collisions,
and express the convoluted cross section in terms of
four observables, $\Sigma_1$ to $\Sigma_4$, which 
contain all the information about the $\gamma\gamma
\rightarrow t\overline{t}$ helicity amplitudes.
When we do not study $W^+$ and $W^-$ decay angular distributions,
the differential cross sections is expressed as
\begin{eqnarray}
\label{diffcross}
&&\frac{d\sigma}{dm_{t\overline{t}}~d\cos\theta~ d\phi~
d\cos\overline{\theta}~ d\overline{\phi}}
\\ \nonumber
&&=
\int d\sqrt{\hat{s}} \sum_{\lambda_1,~\lambda_2} \left(
\frac{1}{{\cal L}_{0.8}} \frac{d{\cal L}^{\lambda_1\lambda_2}}
{d\sqrt{\hat{s}}} \right) \left(
\frac{d\hat{\sigma}_{\lambda_1\lambda_2}
(\sqrt{\hat{s}})}{d\cos\theta~ d\phi~
d\cos\overline{\theta}~ d\overline{\phi}} \right)
G(\sqrt{\hat{s}}-m_{t\overline{t}},\Delta)
\\ \nonumber
&&\equiv \{ \Sigma_1(m_{t\overline{t}})
\left[(B_L^2+B_T^2)(1+\cos\theta \cos\overline{\theta})
     +2B_L B_T(1-\cos\theta \cos\overline{\theta})
     +(B_L^2-B_T^2)(\cos\theta + \cos\overline{\theta}) 
     \right]
\\ \nonumber
&&+ \Sigma_2(m_{t\overline{t}})
\left[(B_L^2+B_T^2)(1+\cos\theta \cos\overline{\theta})
     +2B_L B_T(1-\cos\theta \cos\overline{\theta})
     -(B_L^2-B_T^2)(\cos\theta + \cos\overline{\theta}) 
     \right]
\\ \nonumber
&&+ \Sigma_3(m_{t\overline{t}})
\left[(B_L-B_T)^2 \sin\theta \sin\overline{\theta} 
\cos(\phi-\overline{\phi})\right]
\\ \nonumber
&&+ \Sigma_4(m_{t\overline{t}})
\left[-(B_L-B_T)^2 \sin\theta \sin\overline{\theta} 
\sin(\phi-\overline{\phi})\right] \}/16\pi^2
\\ \nonumber
&&+~~\left[ ( \sigma=-\overline{\sigma})~~ {\rm contributions}\right] .
\end{eqnarray}
Here small non-resonant contributions from 
$\sigma=-\overline{\sigma}$ ($RL$ or $LR$) events are
not shown explicitly.
The four coefficients of the distinct decay angular 
distributions are 
\begin{eqnarray}
\Sigma_i(m_{t\overline{t}})=
\int d\sqrt{\hat{s}} \sum_{\lambda_1,~\lambda_2}
\left( \frac{1}{{\cal L}_{0.8}} 
\frac{d{\cal L}^{\lambda_1\lambda_2}}
{d\sqrt{\hat{s}}} \right) \left( \frac{3\beta}{32\pi \hat{s}}
\int S^i_{\lambda_1 \lambda_2} (\Theta, \sqrt{\hat{s}})d\cos\Theta
\right)
~G(\sqrt{\hat{s}}-m_{t\overline{t}},\Delta),
\\ \nonumber
~~~~~~~~~~~~~~~~~~~~~~~~~~~~~~~~~~~~~~~~~~~~~~~~~{\rm for}~~i=1-4,
\end{eqnarray}
where the functions $S_{\lambda_1 \lambda_2}^i$ contain
all the information about the $\gamma\gamma \rightarrow 
t\overline{t}$ helicity amplitudes:
\begin{eqnarray}
\label{S1toS4}
S^1_{\lambda_1 \lambda_2} &=& \left| {\cal M}_{\lambda_1\lambda_2}^{RR} \right|^2,
\\ \nonumber
S^2_{\lambda_1 \lambda_2} &=& \left|{\cal M}_{\lambda_1\lambda_2}^{LL}\right|^2,
\\ \nonumber
S^3_{\lambda_1 \lambda_2} &=&
2\Re\left[{\cal M}_{\lambda_1\lambda_2}^{RR} \left({\cal M}
_{\lambda_1\lambda_2}^{LL}\right)^*\right],
\\ \nonumber
S^4_{\lambda_1 \lambda_2} &=&
2\Im\left[{\cal M}_{\lambda_1\lambda_2}^{RR} \left({\cal M}
_{\lambda_1\lambda_2}^{LL}\right)^*\right].
\end{eqnarray}
A few remarks about eq.~(\ref{diffcross}) are in order.
The compact expression for the differential cross section in
terms of the observable $m_{t\overline{t}}$, the
$t \rightarrow bW^+$ decay angles $\theta$ and $\phi$,
and the $\overline{t} \rightarrow \overline{b} W^-$
decay angles $\overline{\theta}$ and $\overline{\phi}$ are
obtained by integrating out the 
$\gamma\gamma \rightarrow t\overline{t}$ scattering angle
$\Theta$, the $W^+$ decay angles $\theta^*$ and $\phi^*$,
and the $W^-$ decay angles $\overline{\theta^*}$ and 
$\overline{\phi^*}$; see eq.~(\ref{comdiffcross}). 
We do not lose much information by the integration
over $\cos\Theta$ because the resonant $J=0$ amplitudes
do not depend on $\cos\Theta$ and because the $\cos\Theta$
dependences of the interfering QED amplitudes are mild
near the $t\overline{t}$ threshold; $\beta=0.48$ at
$\sqrt{s}_{\gamma\gamma}=400$ GeV. As explained in
Sec.~3.2, a careful study of $W^+$ and $W^-$ decay
angular distributions should give us independent
measurements of the observables $\Sigma_1$ to $\Sigma_4$,
and should therefore reduce errors.

\begin{figure}[t]
\begin{center}
\epsfxsize=15cm
\epsfysize=10cm
\epsffile{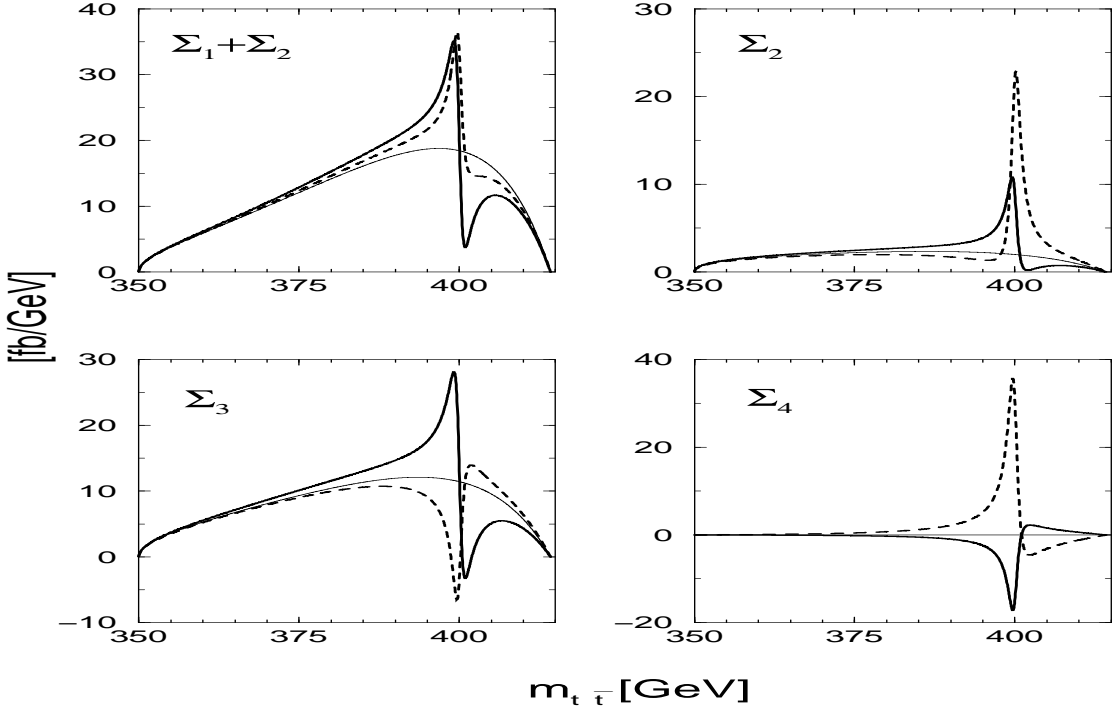}
\caption{The observables $\Sigma_1$ to $\Sigma_4$
with no smearing by detector resolution.
The thick solid (dashed) curves show the predictions for
the $A$ ($H$) production. The thin solid curves show the
QED predictions with no Higgs production.
}\label{tcross09}
\end{center}
\end{figure}

\begin{figure}[t]
\begin{center}
\epsfxsize=15cm
\epsfysize=10cm
\epsffile{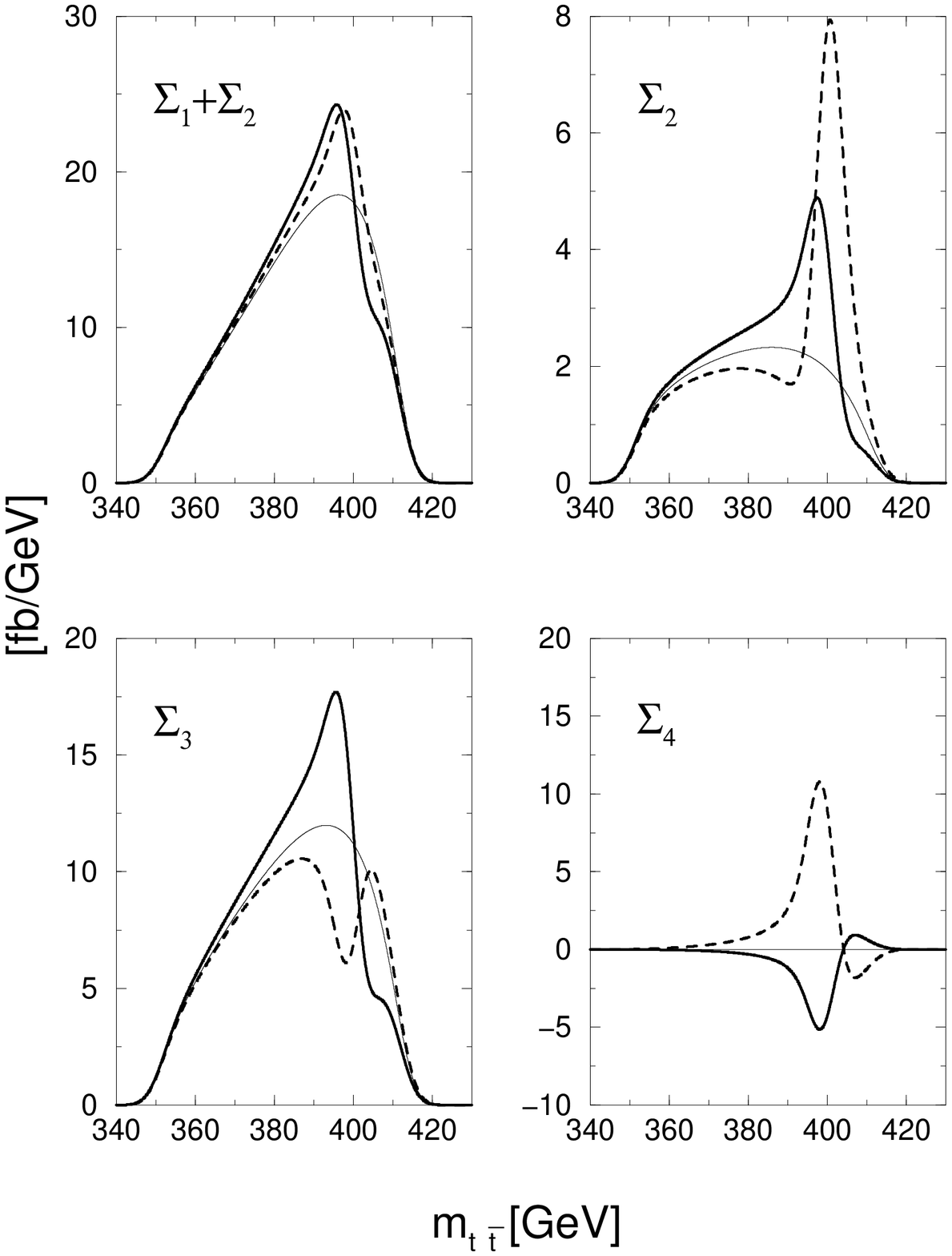}
\caption{The observables $\Sigma_1$ to $\Sigma_4$
with the $t\overline{t}$ invariant mass measurement
resolution factor $\Delta=3$ GeV.
The thick solid (dashed) curves show the predictions for
the $A$ ($H$) production. The thin solid curves show the
QED predictions with no Higgs production.
} \label{tcross39}
\end{center}
\end{figure}

\begin{figure}[t]
\begin{center}
\epsfxsize=15cm
\epsfysize=10cm
\epsffile{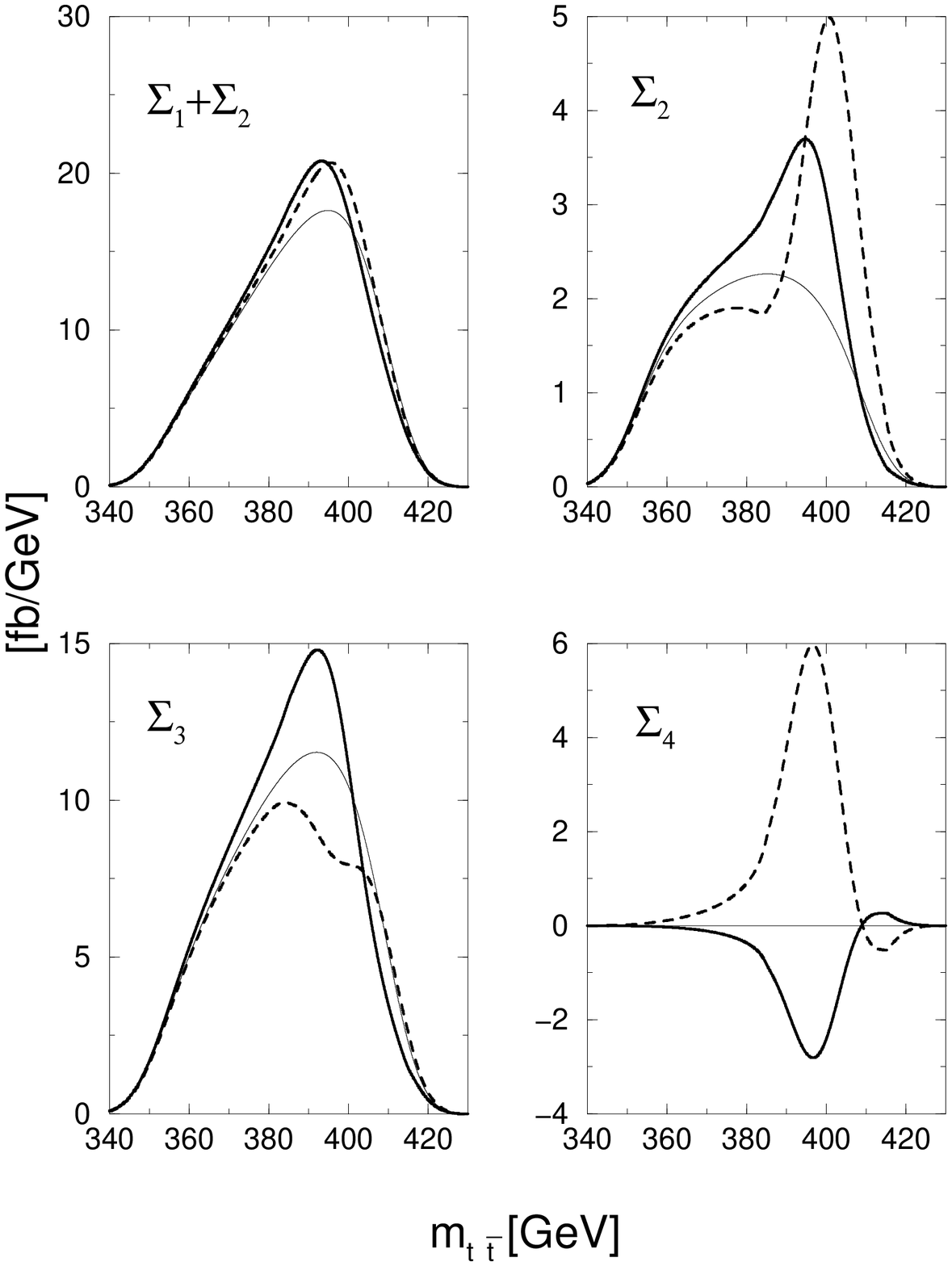}
\caption{The observables $\Sigma_1$ to $\Sigma_4$
with the $t\overline{t}$ invariant mass measurement
resolution factor $\Delta=6$ GeV.
The thick solid (dashed) curves show the predictions for
the $A$ ($H$) production. The thin solid curves show the
QED predictions with no Higgs production.
} \label{tcross69}
\end{center}
\end{figure}

The four observables $\Sigma_1$ to $\Sigma_4$ of 
eq.~(\ref{diffcross}) are shown in Fig.~\ref{tcross09} for
$\Delta=0$ GeV (no smearing by
detector resolution), Fig.~\ref{tcross39} for $\Delta=3$ GeV and 
in Fig.~\ref{tcross69} for $\Delta=6$ GeV. The predictions of
the $A$ and $H$ productions are shown by thick-solid and 
thick-dashed curves, respectively.
The QED predictions are shown by the thin-solid lines.
The quantity
$\Sigma_1 + \Sigma_2$ is simply the total $t\overline{t}$
production cross section, smeared by the resolution factor
of $\Delta$. We show $\Sigma_2$ instead of $\Sigma_1$ because
the $A$ and $H$ production amplitudes interfere with the QED
amplitudes differently in the $\lambda_1=\lambda_2=+$ to
$\sigma=\overline{\sigma}=L$ amplitudes.

When we draw the predictions of $A$ and $H$ productions in
Figs.~\ref{tcross09}, \ref{tcross39} and \ref{tcross69},
we adopt a MSSM prediction for the $A$ production, while
the $H$ production curves are drawn by using the amplitudes
${\cal M}_H$ which are obtained from the ${\cal M}_A$ for
the same mass and width and the same magnitudes for the
partial widths to $\gamma\gamma$ and $t\overline{t}$.
The MSSM parameters used for calculating ${\cal M}_A$
are as follows: $m_A=400$ GeV, $\tan\beta=3$,
$m_{\widetilde{f}}=1$ TeV, $M_2=500$ GeV, $\mu=-500$ GeV.
We find $m_A=400$ GeV, $\Gamma_A=1.75$ GeV,
Br$(A\rightarrow\gamma\gamma)=1.53\times 10^{-5}$ and
Br$(A\rightarrow t\overline{t})$=0.946 for the
above parameters~\cite{HDECAY}. 
The $H$ production amplitudes ${\cal M}_H$
are thus obtained from ${\cal M}_A$ by keeping the mass, 
width and partial widths common in order to show clearly the
sensitivity of the four observables to the CP property of the
produced spinless boson.
For the collider parameters, we use
$E_e=250$ GeV, $P_l=-1.0$, $P_e=0.9$ and $x=4.8$, 
where colliding photons
are highly polarized to be $+$ around $\sqrt{\hat{s}}=400$ GeV;
see Fig.~{\ref{luminosity}}. 
Since the effects from the $(\lambda_1 \lambda_2)=$
($+-$), ($-+$) and ($- -$) 
combinations on the observables are less than 1\% 
around the peak region, they are neglected here.
In this limit, the quantities $S_{1-4}$ in eq.~(\ref{S1toS4})
can be expressed by 
${\cal M}_t$ and ${\cal M}_\phi$ as
\begin{eqnarray}
S_{++}^1 &=& 
\left| {\cal M}_t \right|^2 + \left| {\cal M}_\phi \right|^2
+2{\cal M}_t \Re\left[{\cal M}_\phi \right],
\label{s1} \\ 
S_{++}^2 &=& 
\left(\frac{1-\beta}{1+\beta}\right)^2 \left|{\cal M}_t\right|^2
+\left| {\cal M}_\phi \right|^2 \pm 2 \frac{1-\beta}{1+\beta}
{\cal M}_t \Re \left[ {\cal M}_\phi \right],
\label{s2} \\
S_{++}^3 &=&
2\frac{1-\beta}{1+\beta}\left|{\cal M}_t\right|^2 \pm
2\left|{\cal M}_\phi\right|^2 + 2\left(\frac{1-\beta}{1+\beta}
\pm 1\right){\cal M}_t \Re\left[{\cal M}_\phi\right],
\label{s3} \\
S_{++}^4 &=&
2\left(\frac{1-\beta}{1+\beta}
\mp 1\right){\cal M}_t \Im\left[{\cal M}_\phi\right]
, \label{s4}
\end{eqnarray}
where the upper and lower signs are adopted for $A$ and $H$, 
respectively.

Let us now examine carefully the results shown
in Fig.~\ref{tcross09} to \ref{tcross69}.
For the total production cross section
$\Sigma_1 + \Sigma_2$, it can be clearly observed 
in Fig.~{\ref{tcross09}} that
the $A$ production amplitudes receive stronger constructive
(destructive) interference below (above) the resonance peak
than the $H$ production amplitudes. A sharp dip above
the resonance peak for the $A$ production line-shape may
be considered as a signal of a CP-odd resonance production.
However, the difference between the $A$ and $H$ line shapes
diminishes by smearing. A hint of strong destructive interference
survives in Fig.~{\ref{tcross39}} for the smearing with
$\Delta=3$ GeV, but the difference essentially disappears
in Fig.~{\ref{tcross69}} for $\Delta = 6$ GeV.
The two thick curves for $\Sigma_1 + \Sigma_2$ in 
Fig.~\ref{tcross69} can only tell broad enhancement over
the QED prediction, which may be fitted well by both
$A$ and $H$ production assumptions with slightly different
mass and width values.

The $\Sigma_2$ shows
not only large contribution of the Higgs production
but also the interference effects which have opposite
contribution for the $A$ and $H$ production. 
The magnitudes of the effects
are small because the QED amplitude which interferes with
the Higgs production amplitudes is suppressed by the factor
of $\frac{1-\beta}{1+\beta}$; see Table~\ref{helamp}.
Here the distinctive signature of
the negative interference below the resonance for the $H$
production may survive even for the resolution of 
$\Delta=6$ GeV in Fig.~\ref{tcross69}.

The interference effects we observe in the $\Sigma_3$ 
is larger for $A$ than for $H$ due to 
the factor of $\frac{1-\beta}{1+\beta}\pm 1$ in eq.~(\ref{s3}).
A sharp dip for the $A$ production line-shape and a small excess
for the $H$ produciton line-shape above the resonance peaks
are the effects. The destructive interference effect for $A$
may survive even in Fig.~\ref{tcross69} for $\Delta=6$ GeV,
whereas the small constructive interference effect for $H$
almost disappears in Fig.~\ref{tcross69}.
It is notable that the effects of the Higgs production
has opposite signs for $A$ and $H$ in eq.~(\ref{s3}).
This oppositeness causes that the $A$ production enhances
$\Sigma_3$ above the QED prediction near the peak of the total
cross section $\Sigma_1 + \Sigma_2$, whereas the $H$ 
production predicts smaller $\Sigma_3$ than the QED prediction
around the peak of the cross section. This feature seems to 
persist even with faint $t\overline{t}$ mass resolution,
in Fig.~\ref{tcross39} for $\Delta=3$ GeV and Fig.~\ref{tcross69}
for $\Delta=6$ GeV.

As for the $\Sigma_4$, the pure interference effects can be
observed. The QED amplitudes predict $\Sigma_4$=0 because we adopt the
tree-level amplitudes in our analysis \footnote{The continuum
$\gamma\gamma \rightarrow t\overline{t}$ amplitudes should have
imaginary parts of the order of $\alpha_s$ in QCD perturbation
theory.}. The $A$ production predicts negative and the $H$ 
production predicts positive effects for $\Sigma_4$ around
the production peak.
The difference in the magnitudes comes from  
the factor of $\frac{1-\beta}{1+\beta}\mp 1$ in eq.~(\ref{s4}).
These characteristics appear even considering the detector resolution
as is shown in Fig.~\ref{tcross39} for $\Delta=3$ GeV and 
Fig.~\ref{tcross69} for $\Delta=6$ GeV.
The imaginary part of the interference
term, $\Sigma_4$, discriminates between $A$ and $H$ most clearly.

Summing up, we have made the following observation in this
subsection. The $m_{t\overline{t}}$ dependence of the
total production cross section, $\Sigma_1 + \Sigma_2$,
can in principle reveal the difference between
$A$ and $H$ productions, as shown in Fig.~\ref{tcross09}.
However the distinctive signatures of the $A$ productions,
the constructive interference below the resonance and the
pronounced destructive interference above the resonance
diminish as the $m_{t\overline{t}}$ measurement resolution
becomes worse to $\Delta=3$ GeV (Fig.~\ref{tcross39}) and
to $\Delta=6$ GeV (Fig.~\ref{tcross69}).
It is only the tiny destructive interference effects above
the resonance in Fig.~\ref{tcross69} which signals 
the production of $A$ rather than $H$. The situation
slightly improves by observing the $\Sigma_2$ component
by selecting those events where the produced top-quarks
are both left-handed. Here the distinctive signature of
the negative interference below the resonance for the $H$
production may survive even for the resolution of 
$\Delta=6$ GeV in Fig.~\ref{tcross69}.
The cross section for $t_L \overline{t}_L$ production,
however, is rather small as compared to the dominant
$t_R \overline{t}_R$ production, because of the
$(1-\beta)/(1+\beta)$ suppression factor in the corresponding
QED amplitude; see Table~\ref{helamp}.
Further information are obtained by studying the interference
between the $t_R \overline{t}_R$ and the
$t_L \overline{t}_L$ amplitudes in the observables
$\Sigma_3$ and $\Sigma_4$. The real part of the interference
term, $\Sigma_3$, shows that the $A$ production enhances
$\Sigma_3$ above the QED prediction near the peak of the total
cross section, $\Sigma_1 + \Sigma_2$, whereas the $H$ 
production predicts smaller $\Sigma_3$ than the QED prediction
around the peak of the cross section. This feature seems to 
persist even with faint $t\overline{t}$ mass resolution,
in Fig.~\ref{tcross39} for $\Delta=3$ GeV and Fig.~\ref{tcross69}
for $\Delta=6$ GeV. Finally the imaginary part of the interference
term, $\Sigma_4$, discriminates between $A$ and $H$ most clearly.
The $A$ production predicts negative and the $H$ 
production predicts positive effects for $\Sigma_4$ around
the production peak. We therefore propose to use the four
observables $\Sigma_1$ to $\Sigma_4$ in determining the CP
property of the spin zero resonance in the $\gamma\gamma
\rightarrow t\overline{t}$ channel.

In the above discussion, we studied four observables separately.
Once they are derived individually, we can obtain their
arbitrary linear combinations.
The most powerful combinations for probing the CP parity
of Higgs bosons are $\Sigma_1 + \Sigma_2 + \Sigma_3$ and
$\Sigma_1 + \Sigma_2 - \Sigma_3$.
The former combination receives contribution only from
the CP-odd resonance, while the latter only from 
the CP-even resonance when CP is conserved.
It is therefore straightforward to separate the CP-even and
CP-odd resonances, even when their masses are degenerate.

\subsection{Effects of the 
$\gamma\gamma\phi$ phase on the observables}

In this subsection, we study the $\arg(b_\gamma^\phi)$
dependence of the four observables studied in the previous
subsection. We first re-parameterize the $J_z=0$ amplitudes of
eq.~(\ref{ampeq}) as follows:
\begin{eqnarray}
{\cal M}_{\lambda \lambda}^{\sigma \sigma}=
\left[{\cal M}_t \right]_{\lambda \lambda}^{\sigma \sigma}
+ \left( \frac{\sqrt{\hat{s}}}{m_\phi} \right)^3
r_\phi \cdot i \left[ 1+{\rm exp}\left(
2i \tan^{-1} \frac{s^2-m_\phi^2}{m_\phi \Gamma_\phi} \right)
\right],
\end{eqnarray}
where
$r_H=\sigma \beta b_\gamma^H d_t^H m_H/(2\Gamma_H)$
and $r_A=\lambda b_\gamma^A d_t^A m_A/(2\Gamma_A)$.
In this expression, the phase of the Breit-Wigner resonance
amplitude is shifted by the phase of the $r_\phi$ factor
which is essentially the phase of the $\gamma\gamma\phi$ vertex
factor $b_\gamma^\phi$ if we neglect the phase in the 
$t \overline{t} \phi$ vertex $d_t^\phi$.
It should also be noted that 
\begin{eqnarray}
|r_\phi|^2 = \frac{32 \pi^2}{3 \beta} {\rm Br}
(\phi \rightarrow \gamma\gamma) {\rm Br}
(\phi \rightarrow t\overline{t}).
\end{eqnarray}

In the above discussions, we draw the $H$ production curves
by assuming not only $m_H=m_A$, $\Gamma_H=\Gamma_A$ and
${\rm Br}(H\rightarrow \gamma\gamma) {\rm Br}(H \rightarrow
t\overline{t})={\rm Br}(A\rightarrow \gamma\gamma) {\rm Br}
(A \rightarrow t\overline{t})$, but also that the 
$\gamma\gamma \rightarrow H$ amplitude is proportional to the
$\gamma\gamma \rightarrow A$ amplitude as a complex numbers,
\begin{eqnarray}
b_\gamma^H = b_\gamma^A \left[ \frac{\Gamma(H 
\rightarrow \gamma\gamma)}{\Gamma(A \rightarrow \gamma\gamma)}
\right]^{1/2}.
\end{eqnarray}
We note here that the phase of the $H\rightarrow\gamma\gamma$
amplitude, $\arg(b_\gamma^H)$, and that of the $A \rightarrow
\gamma\gamma$ amplitude, $\arg(b_\gamma^A)$, depend significantly
in the model parameters. As an example, we show in 
Table~\ref{bpara} the MSSM prediction for the real and imaginary 
parts of $b_\gamma^A$ and $b_\gamma^H$.
Here, we calculate the $A$ and $H$ masses and couplings for the 
MSSM parameters; $m_A=400$ GeV, $\tan\beta=3$,
$m_{\widetilde{f}}=1$ TeV, $M_2=500$ GeV, $\mu=-500$ GeV. 
We find that $\arg(b_\gamma^A)$ is much larger than
$\arg(b_\gamma^H)$ . The large imaginary part of 
$b_\gamma^A$ is a result of the $s$-wave $A \rightarrow
t \overline{t}$ decay near the $t\overline{t}$ production
threshold. The imaginary part of $b_\gamma^H$ is
suppressed by the $p$-wave $H \rightarrow t \overline{t}$
decay and also by the partial cancellation due to the $H
\rightarrow W^+ W^-$ contribution. Therefore, 
in the framework of the two Higgs doublet model without 
any new particles which contribute to the vertex significantly, 
the $A$ boson has relatively large phase and
the $H$ boson has tiny phase.
Because the imaginary part
of the $\phi \rightarrow \gamma \gamma$ amplitude is a sum of
the contribution from the $\phi$ decay modes into charged particles
whereas the real part receives contribution from all the charged 
particles, we expect that $\arg(b_\gamma^\phi)$ is a good probe
of heavy charged particles.

\begin{figure}[h]
\begin{center}
\epsfxsize=18cm
\epsfysize=9cm
\epsffile{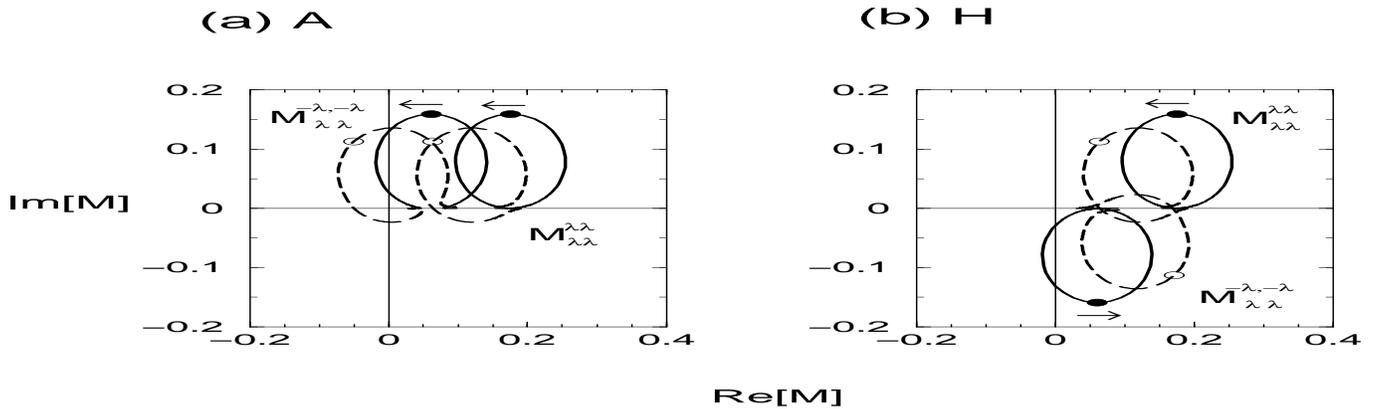}
\caption{ The $\hat{s}$-dependence of the $\gamma\gamma
\rightarrow t\overline{t}$
amplitudes ${\cal M}_{\lambda\lambda}^{\sigma\sigma}$
at $\Theta=0^o$. 
The amplitudes with $A$ production are shown
in the left figure, whereas those for $H$ production
are shown in the right. 
The cases of $arg(b_\gamma^\phi)=0$ and
$\pi/4$ are denoted by
the solid and dashed circles, respectively. 
The small arrows indicate
the direction of increasing $\hat{s}$ and
the solid and open small circles on the trajectories show the
$\hat{s}=m_\phi^2$ points. As $\hat{s}$ grows
the amplitudes make counterclockwise trajectories, and
the magnitude of the resonance amplitude hits its 
maximum at $\hat{s}=m_\phi^2$.}\label{argant}
\end{center}
\end{figure}

Fig.~\ref{argant}
shows plots of the amplitudes ${\cal M}_{\lambda \lambda}
^{\sigma \sigma}$ on the complex plane where the scattering angle
$\Theta$ is fixed to be zero as a sample.
The amplitudes with the $A$ ($H$) production is 
in the left (right) side.
Since the tree amplitudes $\left[{\cal M}_t \right]_{\lambda\lambda}
^{\sigma\sigma}$ are real and almost constant around the resonance,
the plots draw a counterclockwise circle 
which have the beginning- and end-points on the real axis
as $\hat{s}$ increases.
The circles which have the beginning points
nearer (further) from the origin correspond to 
${\cal M}_{\lambda\lambda}
^{-\lambda, -\lambda}$ (${\cal M}_{\lambda \lambda}^{\lambda\lambda}$).  
Two cases of $arg(b_\gamma^\phi)$
are considered. One is the case where
$b_\gamma^\phi$ has no phase (solid curves), the other is
$arg(b_\gamma^\phi)=\pi/4$ (dashed curves).
The solid and open small circles 
on the trajectories indicate the $\hat{s}=m_\phi^2$ points.
When $m_\phi^2-\hat{s} \gg
m_\phi \Gamma_\phi$, the amplitudes are real positive
numbers that are determined by the QED amplitudes of
Table~\ref{treehelamp}.
As $\hat{s}$ grows,
the amplitudes make counterclockwise trajectories, and
the magnitude of the resonance amplitude hits its 
maximum at $\hat{s}=m_\phi^2$.
At $m_\phi^2-\hat{s} \ll
m_\phi \Gamma_\phi$, the amplitudes reduce to the real
and positive QED amplitude again. The trajectories do not
close because of the mild $\hat{s}$-dependence of the
QED amplitudes.
\begin{figure}[h]
\begin{center}
\epsfxsize=15cm
\epsfysize=10cm
\epsffile{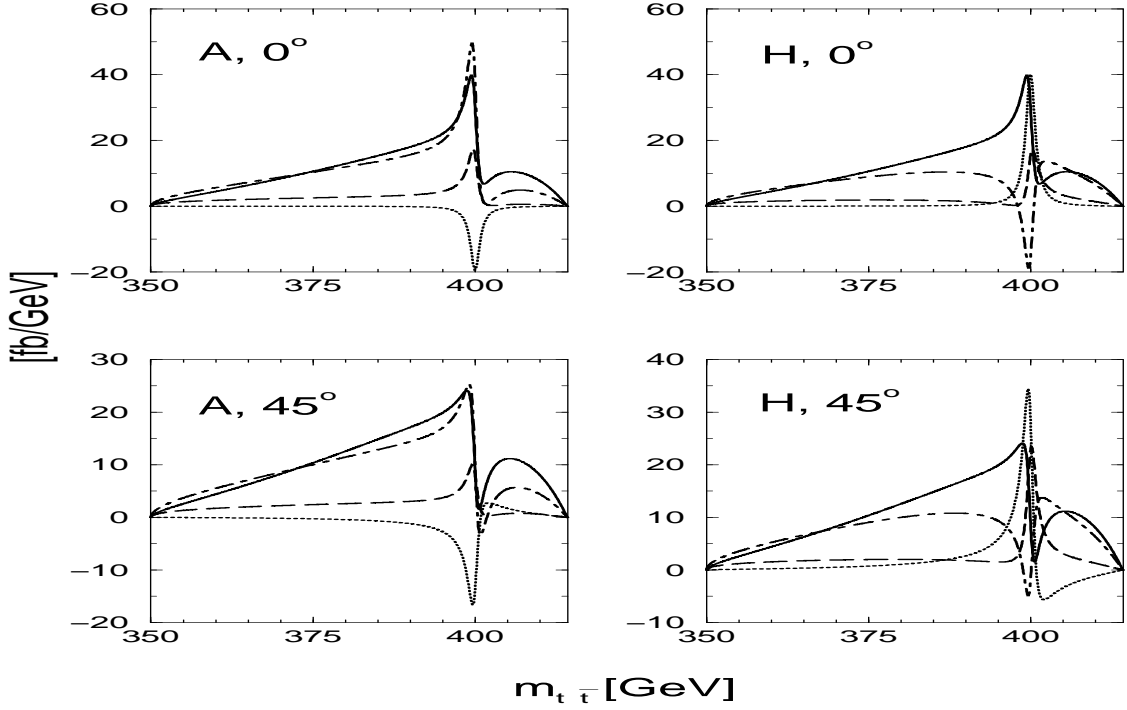}
\caption{The observables $\Sigma_1$ to $\Sigma_4$ 
with no smearing by detector resolution.
The solid, dashed, dot-dashed and dotted curves are
$\Sigma_1$, $\Sigma_2, \Sigma_3$ and $\Sigma_4$, respectively.
The observables with $A$ production are in the left
(right) figures whereas those with the
$H$ production are shown in the right. 
The upper and lower figures 
show the case of
$arg(r_\phi)=0$  and $\pi/4$, respectively. }\label{phase}
\end{center}
\end{figure}

The magnitudes of ${\cal M}_{\lambda\lambda}^{\sigma\sigma}$ 
have peaks at the furthest points
from the origin on the trajectories. 
The $\sqrt{\hat{s}}$ values
at which the amplitudes have the largest magnitude are
almost similar between ${\cal M}_{\lambda\lambda}^{\lambda\lambda}$ 
and ${\cal M}_{\lambda\lambda}^{-\lambda,-\lambda}$
for the $A$ production (slightly below the $\hat{s}=m_A^2$ point),
while they are significantly different
for the $H$ production because the sign of the imaginary parts
are opposite between 
${\cal M}_{\lambda\lambda}^{-\lambda,-\lambda}$ and
${\cal M}_{\lambda\lambda}^{\lambda\lambda}$.
The amplitude of ${\cal M}_{\lambda\lambda}^{\lambda\lambda}$
becomes maximum slightly below the $\hat{s}=m_H^2$ point,
but that of ${\cal M}_{\lambda\lambda}^{-\lambda,-\lambda}$
hits the maximum at $\hat{s} > m_H^2$.

When we compare the $\arg(b_\gamma^A)=0$ amplitudes
(solid circles) and the $\arg(b_\gamma^A)=\pi/4$ amplitudes
(dashed circles), we notice that the magnitudes of all the
amplitudes are reduced for $\arg(b_\gamma^A)>0$ because
the imaginary parts of the resonant amplitudes are positive
for $\arg(b_\gamma^A)=0$. It is notable that at $\hat{s}=m_A^2$
(solid and open circles along the trajectries), the real part of
the ${\cal M}_{\lambda\lambda}^{-\lambda,-\lambda}$ amplitudes 
become negative when $\arg(b_\gamma^A)=\pi/4$. In case of
the $\phi=H$ amplitudes shown in Fig.~\ref{argant}(b),
the most notable feature is that the magnitude of the
${\cal M}_{\lambda\lambda}^{-\lambda,-\lambda}$ amplitudes
increases for $\arg(b_\gamma^A)>0$ because the sign of the 
imaginary part of the $H$ resonant amplitude is negative
for these amplitudes. On the other hand, the magnitudes of
the ${\cal M}_{\lambda\lambda}^{\lambda\lambda}$ amplitudes
decreases for $\arg(b_\gamma^A)>0$ as in the case for the $A$
production amplitudes. 

We show in Fig.~{\ref{phase}} the four observables
$\Sigma_1$ to $\Sigma_4$ for the $A$ production in the left,
and for the $H$ production in the right-hand side.
The predictions for $arg(b_\gamma^\phi)=0$ are shown in the
top figures, whereas those for $arg(b_\gamma^\phi)=\pi/4$
are shown in the bottom figures.

We find that the features which are 
sensitive to the CP parity of the spinless boson $\phi$,
such as the interference pattern of $\Sigma_3$ and $\Sigma_4$
near the resonances, remain stable against varieties of 
$\arg(b_\gamma^\phi)$ between 0 and $\pi/4$.
On the other hand the $\arg(b_\gamma^\phi)$ dependence of
the four observables are significant enough that the phase
of the $\gamma\gamma\phi$ vertex function may be measured
experimentally by a careful study of all the observables. 

\begin{table}[h]
\caption{The values of $b_\gamma^A$ and $b_\gamma^H$.
The loops of $t$, $b$, $W$, $\widetilde{\chi}_1^-$ 
and $\widetilde{\chi}_2^-$ give
large contribution to $b_\gamma^H$ and $b_\gamma^A$
under our parameterization; $m_A=400$ GeV, $\tan\beta=3$,
$M_2=500$ GeV,
$\mu=-500$ GeV and $M_{\widetilde{f}}=1$ TeV. 
$m_H=403.8$ GeV for the above parameters.
\label{bpara}}
\vspace{0.4cm}
\begin{center}
\begin{tabular}{|c|c|c|}
\hline
& $b_\gamma^A\times10^4$ & $b_\gamma^H\times10^4$ \\
\hline
 total & $14+12i$  & $11+1.3i$ \\
\hline\hline
 $t$ & $15+12i$ & $12+3.3i$ \\ 
\hline
 $b$ & $-0.19+0.15i$ & $0.18-0.15i$ \\
\hline
 $W$ & 0.0 & $-1.0-1.7i$ \\
\hline
$\widetilde{\chi}_1^-$ & $-1.1$ & $-1.2$ \\
\hline
$\widetilde{\chi}_2^-$ & $0.51$ & $1.0$ \\
\hline 
\end{tabular}
\end{center}
\end{table}

\section{Conclusions}
We have studied the effects of heavy Higgs bosons in $t\overline{t}$
production process at a PLC. We have introduced observables
which include new type of interference by considering
the angular correlation of decay products of top quarks, and
found that they are useful for probing the CP nature of the
produced Higgs boson.
It has also been shown that variation in the complex phase of the
$\gamma\gamma\phi$ vertex modify the magnitudes of the observables
and the $\sqrt{\hat{s}}$ values where the observables
have peaks and bottoms.

Further studies on the cases where the Higgs sector has
CP non-conservation and/or a degenerate pair of heavy neutral bosons
will be reported elsewhere~\cite{AHfuture}.
The present study may motivate a careful study of the experimental
resolution of the $t\overline{t}$ invariant mass measurements as 
well as a quantitative study on the accuracy of the resonance
parameters, $m_\phi$, $\Gamma_\phi$, Br$(\phi \rightarrow
\gamma\gamma)$Br$(\phi \rightarrow t\overline{t})$,
$\arg(b_\gamma^\phi d_t^\phi)$, and its CP parity.
\vspace{7mm}

{\bf Acknowledgments}
~The authors would like to thank T.~Takahashi and I.~Watanabe 
for useful comments. The work of EA is supported in part by
the Grant-in-Aid for Scientific Research from MEXT, Japan.

\newpage
\appendix
\section{Amplitude for the process $\gamma \gamma 
\rightarrow t \overline{t}
\rightarrow b f_1 \overline{f_2} ~ \overline{b} f_3 \overline{f_4}$}

We describe the helicity
amplitudes for the process $\gamma \gamma \rightarrow
t \overline{t} \rightarrow b f_1 \overline{f_2} ~ 
\overline{b} f_3 \overline{f_4}$
as 
\begin{eqnarray}
&&{\bf M}_{\lambda_1 \lambda_2}
(\Theta; \theta,\phi,\overline{\theta},\overline{\phi};
\theta^*,\phi^*,\overline{\theta}^*,\overline{\phi}^*) 
\\ \nonumber
&&~~~~~~~=
\sum_{\sigma =L,R} \sum_{\Lambda =-,0} \sum_{\overline{\Lambda}=0,+}
{\cal M}_{\lambda_1 \lambda_2}^{\sigma \overline{\sigma}}(\Theta) 
D_{\sigma}^{\Lambda} (\theta,\phi)
\overline{D}_{\overline{\sigma}}^{\overline{\Lambda}}
(\overline{\theta},\overline{\phi})
W_{\Lambda}(\theta^*, \phi^*) \overline{W}_{\overline{\Lambda}}
(\overline{\theta}^*,\overline{\phi}^*),
\end{eqnarray}
in the zero-width limit of the top-quark and the $W$ bosons.
Here 
$\lambda_1$, $\lambda_2$ are the helicities of the colliding photons,
${\cal M}_{\lambda_1 \lambda_2}^{\sigma \overline{\sigma}}(\Theta)$
is the $\gamma(\lambda_1) \gamma(\lambda_2) \rightarrow
t(\sigma) \overline{t}(\overline{\sigma})$ scattering amplitudes
at the scattering angle $\Theta$ in the $\gamma\gamma$ collision
c.o.m.~frame,
$D_{\sigma}^{\Lambda}$ and 
$\overline{D}_{\overline{\sigma}}^{\overline{\Lambda}}$
are the decay amplitudes
for the processes $t_{\sigma} \rightarrow b W^+_{\Lambda}$ 
and $\overline{t}_{\overline{\sigma}} \rightarrow
\overline{b} W^-_{\overline{\Lambda}}$ in the $t$ and $\overline{t}$ 
rest frame, respectively.
$W_{\Lambda}$ and $\overline{W}_{\overline{\Lambda}}$
are the decay amplitudes for  
the processes $W^+_\Lambda \rightarrow f_1 \overline{f_2}$ 
and $W^-_{\overline{\Lambda}}
 \rightarrow f_3 \overline{f_4}$ in the decaying $W$ rest frames,
in the massless fermion limit ($m_{f_i}=0$).
The decay amplitudes have the following simple forms in the phase
connection of Ref.~\cite{HZ1986, HELAS}:
\begin{center}
\begin{tabular}{llll}
$D_L^0 = \sqrt{\frac{B_L}{2\pi}} \sin\frac{\theta}{2}$,
&
$D_L^- = \sqrt{\frac{B_T}{2\pi}} \cos\frac{\theta}{2}$,
&
$D_R^0 = \sqrt{\frac{B_L}{2\pi}} \cos\frac{\theta}{2} e^{i\phi}$,
&
$D_R^- = -\sqrt{\frac{B_T}{2\pi}} \sin\frac{\theta}{2} e^{i\phi}$,
\\
$\overline{D}_L^0 = -\sqrt{\frac{B_L}{2\pi}} \sin\frac{\overline{\theta}}{2}$,
&
$\overline{D}_L^+ = \sqrt{\frac{B_T}{2\pi}} \cos\frac{\overline{\theta}}{2}$,
&
$\overline{D}_R^0 = -\sqrt{\frac{B_L}{2\pi}} \cos\frac{\overline{\theta}}{2}
e^{-i \overline{\phi}}$,
&
$\overline{D}_R^+ = -\sqrt{\frac{B_T}{2\pi}} \sin\frac{\overline{\theta}}{2}
e^{-i \overline{\phi}}$,
\end{tabular}
\end{center}
\begin{eqnarray}
\end{eqnarray}
and
\begin{center}
\begin{tabular}{ll}
$W_0 = \sqrt{\frac{3}{8\pi} B_{12}} \sin \theta^*$,
&
$W_- = \sqrt{\frac{3}{8\pi} B_{12}}
    \frac{1- \cos \theta^*}{\sqrt{2}} e^{-i\phi^*}$,
\\
$\overline{W}_0 = \sqrt{\frac{3}{8\pi} B_{34}}
    \sin \overline{\theta}^*$,
&
$\overline{W}_+ = -\sqrt{\frac{3}{8\pi} B_{34}}
    \frac{1- \cos \overline{\theta}^*}{\sqrt{2}} 
e^{i\overline{\phi}^*}$.
\end{tabular}
\end{center}
Here the decay amplitudes are normalized as
\begin{eqnarray}
\int |D_\sigma^-|^2 d\cos\theta d\phi = \int 
|\overline{D}_{\overline{\sigma}}^+|^2 
d\cos\overline{\theta} d\overline{\phi} = B_T =
\frac{2m_W^2}{m_t^2+2m_W^2},
\\ \nonumber
\int |D_\sigma^0|^2 d\cos\theta d\phi
=\int |\overline{D}_{\overline{\sigma}}^0|^2 
d\cos\overline{\theta} d\overline{\phi} = B_L =
\frac{m_t^2}{m_t^2+2m_W^2},
\end{eqnarray}
and 
\begin{eqnarray}
\int |W_\Lambda|^2 d\cos\theta^* d\phi^* = B_{12},
\\ \nonumber
\int |\overline{W}_{\overline{\Lambda}}|^2 
d\cos\overline{\theta}^* d\overline{\phi}^* = B_{34},
\end{eqnarray}
where $B_{12}$ is the branching fraction of $W^+ \rightarrow
f_1 \overline{f}_2$ decays, and $B_{34}$ is that of
$W^- \rightarrow f_3 \overline{f}_4$.
The angles 
$\theta$ and $\phi$ ($\overline{\theta}$ and $\overline{\phi}$)
are, respectively, the polar and azimuthal angles of $W^+$ ($W^-$)
in the $t$ ($\overline{t}$) rest frame where the common polar axis
is chosen along the $t$-momentum direction in the $\gamma\gamma$
collision c.m. frame, and the azimuthal angles $\phi$ and
$\overline{\phi}$ are measured from the $\gamma\gamma \rightarrow
t\overline{t}$ scattering plane.
$\theta^*$ and $\phi^*$ are, respectively,
the polar and azimuthal angles of $\overline{f}_2$
in the $W^+ \rightarrow f_1 \overline{f}_2$ decay rest frame,
whereas $\overline{\theta}^*$ and $\overline{\phi}^*$ are
those of $f_3$ in the $W^- \rightarrow f_3 \overline{f}_4$
rest frame. We choose the $\overline{f}_2$ and $f_3$ momenta
in the above decays so that the angles are those of the
charged leptons in the decays $W^+ \rightarrow \nu_l l^+$
and $W^- \rightarrow l^- \overline{\nu}_l$.
The polar axis are chosen along the $W^\pm$ momentum in the
parent $t$ or $\overline{t}$ rest frame, while the azimuthal
angles $\phi^*$ and $\overline{\phi}^*$ are measured from the
$t \rightarrow b W^+$ and $\overline{t} \rightarrow \overline{b}
W^-$ decay planes, respectively, in the $\gamma\gamma$
collision c.m. frame.

The amplitudes (A.1) can now be expressed solely
in terms of the $\gamma\gamma \rightarrow t\overline{t}$ amplitudes 
${\cal M}_{\lambda_1 \lambda_2}^{\sigma \overline{\sigma}}(\Theta)$:
\begin{eqnarray}
&~&{\bf M}_{\lambda_1 \lambda_2}
(\Theta; \theta,\phi,\overline{\theta},\overline{\phi};
\theta^*,\phi^*,\overline{\theta}^*,\overline{\phi}^*)
/\left(\frac{3}{16\pi^2} \sqrt{B_{12} B_{34}}\right)
\\ \nonumber
&=& {\cal M}_{\lambda_1 \lambda_2}^{LL}(\Theta)
{\Huge \{ } - B_L \sin\frac{\theta}{2}
\sin\frac{\overline{\theta}}{2} \sin\theta^* 
\sin\overline{\theta}^*
\\ \nonumber
&~& ~~~~~~~~~~~~
- \sqrt{B_L B_T} \cos\frac{\theta}{2} 
\sin\frac{\overline{\theta}}{2} 
\frac{1-\cos\theta^*}{\sqrt{2}} \sin\overline{\theta}^*
e^{-i\phi^*} 
\\ \nonumber
&~& ~~~~~~~~~~~~ - \sqrt{B_L B_T} \sin\frac{\theta}{2}
\cos\frac{\overline{\theta}}{2}
\sin\theta^* \frac{1-\cos\overline{\theta}^*}{\sqrt{2}} 
e^{i \overline{\phi}^*}
\\ \nonumber
&~& ~~~~~~~~~~~~
- B_T \cos\frac{\theta}{2} \cos\frac{\overline{\theta}}{2} 
\frac{(1-\cos\theta^*)(1-\cos\overline{\theta}^*)}{2}
e^{-i( \phi^* - \overline{\phi}^* )}  {\Huge \} }
\\ \nonumber
&+& {\cal M}_{\lambda_1 \lambda_2}^{RR}(\Theta)
{\Huge \{ } - B_L \cos\frac{\theta}{2}
\cos\frac{\overline{\theta}}{2} e^{i (\phi - \overline{\phi})} 
\sin\theta^* \sin\overline{\theta}^*
\\ \nonumber
&~& ~~~~~~~~~~~~
+ \sqrt{B_L B_T} \sin\frac{\theta}{2} 
\cos\frac{\overline{\theta}}{2} 
e^{i ( \phi - \overline{\phi} ) }
\frac{1-\cos\theta^*}{\sqrt{2}} \sin\overline{\theta}^*
e^{-i\phi^*} 
\\ \nonumber
&~& ~~~~~~~~~~~~ + \sqrt{B_L B_T} \cos\frac{\theta}{2}
\sin\frac{\overline{\theta}}{2} e^{i (\phi - \overline{\phi})}
\sin\theta^* \frac{1-\cos\overline{\theta}^*}{\sqrt{2}} 
e^{i \overline{\phi}^*}
\\ \nonumber
&~& ~~~~~~~~~~~~
- B_T \sin\frac{\theta}{2} \sin\frac{\overline{\theta}}{2} 
e^{i ( \phi - \overline{\phi} ) }
\frac{(1-\cos\theta^*)(1-\cos\overline{\theta}^*)}{2}
e^{-i( \phi^* - \overline{\phi}^* )}  {\Huge \} } 
\\ \nonumber
&+& {\cal M}_{\lambda_1 \lambda_2}^{LR}(\Theta)
{\Huge \{ } - B_L \sin\frac{\theta}{2}
\cos\frac{\overline{\theta}}{2} e^{-i \overline{\phi}}
\sin\theta^* \sin\overline{\theta}^*
\\ \nonumber
&~& ~~~~~~~~~~~~
- \sqrt{B_L B_T} \cos\frac{\theta}{2} 
\cos\frac{\overline{\theta}}{2}
e^{-i \overline{\phi}} 
\frac{1-\cos\theta^*}{\sqrt{2}} \sin\overline{\theta}^*
e^{-i\phi^*} 
\\ \nonumber
&~& ~~~~~~~~~~~~ + \sqrt{B_L B_T} \sin\frac{\theta}{2}
\sin\frac{\overline{\theta}}{2} e^{-i \overline{\phi}}
\sin\theta^* \frac{1-\cos\overline{\theta}^*}{\sqrt{2}} 
e^{i \overline{\phi}^*}
\\ \nonumber
&~& ~~~~~~~~~~~~
+ B_T \cos\frac{\theta}{2} \sin\frac{\overline{\theta}}{2} 
e^{-i \overline{\phi}}
\frac{(1-\cos\theta^*)(1-\cos\overline{\theta}^*)}{2}
e^{-i( \phi^* - \overline{\phi}^* )}  {\Huge \} }
\\ \nonumber
&+& {\cal M}_{\lambda_1 \lambda_2}^{RL}(\Theta)
{\Huge \{ } - B_L \cos\frac{\theta}{2}
\sin\frac{\overline{\theta}}{2} e^{i \phi } 
\sin\theta^* \sin\overline{\theta}^*
\\ \nonumber
&~& ~~~~~~~~~~~~
+ \sqrt{B_L B_T} \sin\frac{\theta}{2} 
\sin\frac{\overline{\theta}}{2} 
e^{i \phi}
\frac{1-\cos\theta^*}{\sqrt{2}} \sin\overline{\theta}^*
e^{-i\phi^*} 
\\ \nonumber
&~& ~~~~~~~~~~~~ - \sqrt{B_L B_T} \cos\frac{\theta}{2}
\cos\frac{\overline{\theta}}{2} e^{i \phi}
\sin\theta^* \frac{1-\cos\overline{\theta}^*}{\sqrt{2}} 
e^{i \overline{\phi}^*}
\\ \nonumber
&~& ~~~~~~~~~~~~
+ B_T \sin\frac{\theta}{2} \cos\frac{\overline{\theta}}{2} 
e^{i \phi}
\frac{(1-\cos\theta^*)(1-\cos\overline{\theta}^*)}{2}
e^{-i( \phi^* - \overline{\phi}^* )}  {\Huge \} }.
\end{eqnarray}
The differential cross section of eq.~(\ref{comdiffcross})
is now expressed in terms of the $\gamma\gamma \rightarrow
t \overline{t}$ amplitudes
${\cal M}_{\lambda_1 \lambda_2}^{\sigma \overline{\sigma}}(\Theta)$.

\section{Cross section for the process $\gamma \gamma
\rightarrow t \overline{t} \rightarrow bW^+ \overline{b}W^-$}
By using the $W^+$ and $W^-$ decay angular distributions of
eq.~(\ref{comdiffcross})
and the appendix A, one can project out the polarized
$W^+ W^-$ production cross sections.
The cross section for the process $\gamma(\lambda_1) 
\gamma(\lambda_2)
\rightarrow t \overline{t} \rightarrow bW^+(\Lambda)
\overline{b}W^-(\overline{\Lambda})$
is expressed as
\begin{eqnarray}
\frac{d \hat{\sigma}_{\lambda_1 \lambda_2}}
{d\cos\Theta d\cos\theta d\phi d\cos\overline{\theta} 
d\overline{\phi}}= \frac{\beta}{32\pi \hat{s}}
\sum_{\Lambda=0,-} \sum_{\overline{\Lambda}=0,+}
\left| {\bf M}_{\lambda_1 \lambda_2}
^{\Lambda \overline{\Lambda}}
(\Theta; \theta,\phi,\overline{\theta},\overline{\phi})
\right|^2,
\end{eqnarray}
where
\begin{eqnarray}
\left| {\bf M}_{\lambda_1 \lambda_2}
^{\Lambda \overline{\Lambda}}
(\Theta; \theta,\phi,\overline{\theta},\overline{\phi})
\right|^2
&=& \left| {\cal M}_{\lambda_1 \lambda_2}^{RR} \right|^2
\left| D_R^\Lambda \right|^2
\left| \overline{D}_R^{\overline{\Lambda}} \right|^2 
\\ \nonumber
&+& \left| {\cal M}_{\lambda_1 \lambda_2}^{LL} \right|^2
\left| D_L^\Lambda \right|^2
\left| \overline{D}_L^{\overline{\Lambda}} \right|^2
\\ \nonumber
&+& \left| {\cal M}_{\lambda_1 \lambda_2}^{RL} \right|^2
\left| D_R^\Lambda \right|^2
\left| \overline{D}_L^{\overline{\Lambda}} \right|^2
\\ \nonumber
&+& \left| {\cal M}_{\lambda_1 \lambda_2}^{LR} \right|^2
\left| D_L^\Lambda \right|^2
\left| \overline{D}_R^{\overline{\Lambda}} \right|^2
\\ \nonumber
&+& 2 \Re \left[ {\cal M}_{\lambda_1 \lambda_2}^{RR} \cdot 
{\cal M}_{\lambda_1 \lambda_2}^{LL*} \right]
\Re \left[ D_R^\Lambda \overline{D}_R^{\overline{\Lambda}}
D_L^{\Lambda*} \overline{D}_L^{\overline{\Lambda}*} \right]
\\ \nonumber
&-& 2 \Im \left[ {\cal M}_{\lambda_1 \lambda_2}^{RR} \cdot
{\cal M}_{\lambda_1 \lambda_2}^{LL*} \right]
\Im \left[ D_R^\Lambda \overline{D}_R^{\overline{\Lambda}}
D_L^{\Lambda*} \overline{D}_L^{\overline{\Lambda}*} \right]
\\ \nonumber
&+& 2 \Re \left[ {\cal M}_{\lambda_1 \lambda_2}^{RR} \cdot 
{\cal M}_{\lambda_1 \lambda_2}^{RL*} \right]
\Re \left[ D_R^\Lambda \overline{D}_R^{\overline{\Lambda}}
D_R^{\Lambda*} \overline{D}_L^{\overline{\Lambda}*} \right]
\\ \nonumber
&-& 2 \Im \left[ {\cal M}_{\lambda_1 \lambda_2}^{RR} \cdot
{\cal M}_{\lambda_1 \lambda_2}^{RL*} \right]
\Im \left[ D_R^\Lambda \overline{D}_R^{\overline{\Lambda}}
D_R^{\Lambda*} \overline{D}_L^{\overline{\Lambda}*} \right]
\\ \nonumber
&+& 2 \Re \left[ {\cal M}_{\lambda_1 \lambda_2}^{RR} \cdot 
{\cal M}_{\lambda_1 \lambda_2}^{LR*} \right]
\Re \left[ D_R^\Lambda \overline{D}_R^{\overline{\Lambda}}
D_L^{\Lambda*} \overline{D}_R^{\overline{\Lambda}*} \right]
\\ \nonumber
&-& 2 \Im \left[ {\cal M}_{\lambda_1 \lambda_2}^{RR} \cdot
{\cal M}_{\lambda_1 \lambda_2}^{LR*} \right]
\Im \left[ D_R^\Lambda \overline{D}_R^{\overline{\Lambda}}
D_L^{\Lambda*} \overline{D}_R^{\overline{\Lambda}*} \right]
\\ \nonumber
&+& 2 \Re \left[ {\cal M}_{\lambda_1 \lambda_2}^{LL} \cdot 
{\cal M}_{\lambda_1 \lambda_2}^{RL*} \right]
\Re \left[ D_L^\Lambda \overline{D}_L^{\overline{\Lambda}}
D_R^{\Lambda*} \overline{D}_L^{\overline{\Lambda}*} \right]
\\ \nonumber
&-& 2 \Im \left[ {\cal M}_{\lambda_1 \lambda_2}^{LL} \cdot
{\cal M}_{\lambda_1 \lambda_2}^{RL*} \right]
\Im \left[ D_L^\Lambda \overline{D}_L^{\overline{\Lambda}}
D_R^{\Lambda*} \overline{D}_L^{\overline{\Lambda}*} \right]
\\ \nonumber
&+& 2 \Re \left[ {\cal M}_{\lambda_1 \lambda_2}^{LL} \cdot 
{\cal M}_{\lambda_1 \lambda_2}^{LR*} \right]
\Re \left[ D_L^\Lambda \overline{D}_L^{\overline{\Lambda}}
D_L^{\Lambda*} \overline{D}_R^{\overline{\Lambda}*} \right]
\\ \nonumber
&-& 2 \Im \left[ {\cal M}_{\lambda_1 \lambda_2}^{LL} \cdot
{\cal M}_{\lambda_1 \lambda_2}^{LR*} \right]
\Im \left[ D_L^\Lambda \overline{D}_L^{\overline{\Lambda}}
D_L^{\Lambda*} \overline{D}_R^{\overline{\Lambda}*} \right]
\\ \nonumber
&+& 2 \Re \left[ {\cal M}_{\lambda_1 \lambda_2}^{RL} \cdot 
{\cal M}_{\lambda_1 \lambda_2}^{LR*} \right]
\Re \left[ D_R^\Lambda \overline{D}_L^{\overline{\Lambda}}
D_L^{\Lambda*} \overline{D}_R^{\overline{\Lambda}*} \right]
.
\end{eqnarray}
It is helpful to write down the squared amplitudes
in the case where $\lambda_1 = \lambda_2 = \lambda$,
because high luminosity and high degree of $\lambda_1=
\lambda_2=\lambda$ polarization for energetic two photon
pairs can be achieved at a PLC by choosing a right combination
of the laser and the $e^-$ beam polarizations.
We find
\begin{eqnarray}
|{\bf M}_{\lambda \lambda}^{00}|^2
          &=& \{ |{\cal M}_{\lambda \lambda}^{RR}|^2 
\cos^2\frac{\theta}{2} 
                                \cos^2\frac{\overline{\theta}}{2}
           +|{\cal M}_{\lambda \lambda}^{LL}|^2 \sin^2\frac{\theta}{2} 
           \sin^2\frac{\overline{\theta}}{2}
\\ \nonumber
          &~&~~+ \frac{1}{2} \Re[{\cal M}_{\lambda \lambda}^{RR} 
\cdot {\cal M}_{\lambda \lambda}^{LL*}]
           \sin\theta \sin\overline{\theta} \cos(\phi-\overline{\phi})
\\ \nonumber
          &~&~~- \frac{1}{2} \Im[{\cal M}_{\lambda \lambda}^{RR} 
\cdot {\cal M}_{\lambda \lambda}^{LL*}]
           \sin\theta \sin\overline{\theta} 
\sin(\phi-\overline{\phi}) \}
          \times \frac{B_L^2}{4\pi^2},
\\ \nonumber
|{\bf M}_{\lambda \lambda}^{0+}|^2
          &=& \{ |{\cal M}_{\lambda \lambda}^{RR}|^2 
\cos^2\frac{\theta}{2} 
                                 \sin^2\frac{\overline{\theta}}{2}
           +|{\cal M}_{\lambda \lambda}^{LL}|^2 
\sin^2\frac{\theta}{2} 
           \cos^2\frac{\overline{\theta}}{2}
\\ \nonumber
          &~&~~- \frac{1}{2} \Re[{\cal M}_{\lambda \lambda}^{RR} 
\cdot {\cal M}_{\lambda \lambda}^{LL*}]
           \sin\theta \sin\overline{\theta} \cos(\phi-\overline{\phi})
\\ \nonumber
          &~&~~+ \frac{1}{2} \Im[{\cal M}_{\lambda \lambda}^{RR} 
\cdot {\cal M}_{\lambda \lambda}^{LL*}]
          \sin\theta \sin\overline{\theta} 
\sin(\phi-\overline{\phi}) \}
          \times \frac{B_L B_T}{4\pi^2},
\\ \nonumber
|{\bf M}_{\lambda \lambda}^{-0}|^2
          &=& \{ |{\cal M}_{\lambda \lambda}^{RR}|^2 
\sin^2\frac{\theta}{2} 
                                 \cos^2\frac{\overline{\theta}}{2}
           +|{\cal M}_{\lambda \lambda}^{LL}|^2 
\cos^2\frac{\theta}{2} 
           \sin^2\frac{\overline{\theta}}{2}
\\ \nonumber
          &~&~~- \frac{1}{2} \Re[{\cal M}_{\lambda \lambda}^{RR} 
\cdot {\cal M}_{\lambda \lambda}^{LL*}]
           \sin\theta \sin\overline{\theta} \cos(\phi-\overline{\phi})
\\ \nonumber
          &~&~~+ \frac{1}{2} \Im[{\cal M}_{\lambda \lambda}^{RR} 
\cdot {\cal M}_{\lambda \lambda}^{LL*}]
          \sin\theta \sin\overline{\theta} 
\sin(\phi-\overline{\phi}) \}
          \times \frac{B_L B_T}{4\pi^2},
\\ \nonumber
|{\bf M}_{\lambda \lambda}^{-+}|^2
          &=& \{ |{\cal M}_{\lambda \lambda}^{RR}|^2 
\sin^2\frac{\theta}{2} 
                                 \sin^2\frac{\overline{\theta}}{2}
           +|{\cal M}_{\lambda \lambda}^{LL}|^2 
\cos^2\frac{\theta}{2} \cos^2\frac{\overline{\theta}}{2}
\\ \nonumber
          &~&~~+ \frac{1}{2} \Re[{\cal M}_{\lambda \lambda}^{RR} 
\cdot {\cal M}_{\lambda \lambda}^{LL*}]
           \sin\theta \sin\overline{\theta} \cos(\phi-\overline{\phi})
\\ \nonumber
          &~&~~- \frac{1}{2} \Im[{\cal M}_{\lambda \lambda}^{RR} 
\cdot {\cal M}_{\lambda \lambda}^{LL*}]
          \sin\theta \sin\overline{\theta} 
\sin(\phi-\overline{\phi}) \}
          \times \frac{B_T^2}{4\pi^2}.
\end{eqnarray}

\newpage
%
%
%
\newcommand{\Journal}[4]{{\sl #1} {\bf #2} {(#3)} {#4}}
\newcommand{\PL}{\sl Phys. Lett.}
\newcommand{\PR}{\sl Phys. Rev.}
\newcommand{\PRL}{\sl Phys. Rev. Lett.}
\newcommand{\NP}{\sl Nucl. Phys.}
\newcommand{\ZP}{\sl Z. Phys.}
\newcommand{\PTP}{\sl Prog. Theor. Phys.}
\newcommand{\NC}{\sl Nuovo Cimento}
\newcommand{\MPL}{\sl Mod. Phys. Lett.}
\newcommand{\PRep}{\sl Phys. Rep.}
\newcommand{\EPJ}{\sl Eur. Phys. J.}
\newcommand{\CPCR}{\sl Comm. Phys. Commun. Res.}
\newcommand{\NEM}{\sl Nucl. Instrum. Meth.}


\begin{thebibliography}{99}
%
\bibitem{JLC}
K.~Abe et al., [ACFA Linear Collider Working Group Collaboration],
`Particle Physics Experiments
at JLC', hep-ph/0109166.
%
\bibitem{TESLA}
J.~Aguilar-Saavedra et al., [ECFA/DESY LC Physics Working Group
Collaboration], 
`Physics at an $e^+ e^-$ Linear Collider',hep-ph/0106315.
%
\bibitem{NLC}
T.~Abe et al., [American Linear Collider Working Group 
Collaboration], in {\sl Proc. of the APS/DPF/DPB Summer Study on
the Future of Particle Physics (Snowmass 2001)} ed. 
R.~Davidson and C.~Quigg, SLAC-R-570 {\sl Resource book for
Snowmass 2001, 30 Jun - 21 Jul 2001, Snowmass, Colorado}.
%
\bibitem{polarization}
    I.F.~Ginzburg, G.L.~Kotkin, S.L.~Panfil, V.G.~Selbo 
    and V.I.~Telnov,
        {\sl Nucl. Instrum. Methods Phys. Res.} {\bf A219},
        5 (1984);\\
    V.I.~Telnov, {\it ibid.} {\bf 294}, 72 (1990).
%
\bibitem{MM}
    M.M.~Muhlleitner, M.~Kramer, M.~Spira and P.M.~Zerwas,
        {\PL} {\bf B508}, 311 (2001);\\
    D.M.~Asner, J.B.~Gronberg and J.F.~Gunion,
        {\PR} {\bf D67}, 035009 (2003).
%
\bibitem{ggwidth}
    G.V.~Jikia,
	{\NP} {\bf B405}, 24 (1993);\\
    M.S.~Berger,
	{\PR} {\bf D48}, 5121 (1993);\\
    T.~Ohgaki, T.~Takahashi and I.~Watanabe,
	{\PR} {\bf D56}, 1723 (1997);\\
    G.~Jikia and S.~Soldner-Rembold,
        {\NEM} {\bf A472}, 133 (2001);\\
    P.~Niezurawski, A.F.~Zarnecki and M.~Krawczyk,
        hep-ph/0207294.
%
\bibitem{Gunion}
    B.~Grazadkowski and J.F.~Ginion,
        {\PL} {\bf B294}, 361 (1992);\\
    M.~Kramer, J.~K\"{u}hn, M.L.~Stong and P.M.~Zerwas,
        {\ZP} {\bf C64}, 21 (1994);\\
    J.F.~Gunion and J.G.~Kelly,
        {\PL} {\bf B333}, 110 (1994).
%
\bibitem{DESYtalk}
    K.~Hagiwara,
	{\sl Nucl. Instrum. Meth.} {\bf A472}, 12 (2001).
%
%
\bibitem{AKSW}
    E.~Asakawa, J.~Kamoshita, A.~Sugamoto and I.~Watanabe,
        {\EPJ} {\bf C14}, 335 (2000).

\bibitem{ACHL}
    E.~Asakawa, S.Y.~Choi, K.~Hagiwara and J.S.~Lee,
        {\PR} {\bf D62}, 115005 (2000).

\bibitem{GRS}
    R.M.~Godbole, S.D.~Rindani and R.K.~Singh,
	{\PR} {\bf D67}, 095009 (2003).
%
\bibitem{HZ1986}
    K.~Hagiwara and D.~Zeppenfeld,
        {\NP} {\bf B274}, 1 (1986).
%
\bibitem{HHG}
    J.F.~Gunion, H.E.~Haber, G.~Kane, S.~Dawson,
        Higgs hunter's guide (Addison-Wesley Publishing Company 1990)
        , and references therein.
%
\bibitem{AHfuture}
    E.~Asakawa and K.~Hagiwara, in preparation
%
\bibitem{HMW}
    K.~Hagiwara, H.~Murayama and I.~Watanabe,
        {\NP} {\bf B367}, 257 (1991).
%
\bibitem{lepton}
    M.~Je\.{z}abek and J.H.~K\"{u}hn,
        {\NP} {\bf B320}, 20 (1989);\\
    G.~Mahlon and S.~Parke,
        {\PR} {\bf D53}, 4886 (1996).
%
\bibitem{NGHIKK}
    M.~Diehl and O.~Nachtman,
	{\ZP} {\bf C62}, 397 (1994);\\
    J.F.~Gunion, B.~Grzadkowski and X.-G.~He,
	{\PRL} {\bf 77}, 5172 (1996);\\
    K.~Hagiwara, S.~Ishihara, J.~Kamoshita and B.A.~Kniehl,
	{\EPJ} {\bf C14}, 457 (2000).
%
\bibitem{HDECAY}
        A.~Djouadi, J.~Kalinowski and M.~Spira,
                {\CPCR} {\bf 108}, 56 (1998).
%
\bibitem{HELAS}
    H.~Murayama, I.~Watanabe and K.~Hagiwara,
        HELAS: HELicity Amplitude Subroutines for
               Feynman Diagram Evaluations,
        KEK Report 91-11 (1992).
\end{thebibliography}
\end{document}